\documentclass[journal ]{new-aiaa}
\usepackage[utf8]{inputenc}
\usepackage{textcomp}

\usepackage{graphicx}
\usepackage{amsmath}

\usepackage{amssymb}
\usepackage[version=4]{mhchem}
\usepackage{siunitx}
\usepackage{longtable,tabularx}
\usepackage{bm}
\usepackage{comment}
\usepackage{algorithm}
\usepackage{algpseudocode}
\usepackage{xcolor}

\usepackage{float}
\usepackage{subcaption}

\algnewcommand{\Input}{\State \textbf{Input:}\ }
\algnewcommand{\Output}{\State \textbf{Output:}\ }
\newcommand{\algcomment}[1]{%
  \Statex \textcolor{gray}{\(\triangleright\) \textit{#1}}%
}
\algrenewcommand\algorithmicthen{\ \textbf{then}}
\algrenewcommand\algorithmicdo{\ \textbf{do}}
\algrenewcommand\algorithmicend{\textbf{end}}
\algrenewcommand\algorithmicwhile{\textbf{while}}
\algrenewcommand\algorithmicfor{\textbf{for}}
\algrenewcommand\algorithmicif{\textbf{if}}
\newcounter{problem}
\newenvironment{problem}[1][]{%
  \refstepcounter{problem}%
  \par\medskip\noindent\textbf{Problem~\theproblem}%
  \if\relax\detokenize{#1}\relax\else\textbf{ (#1)}\fi
  \textbf{.}\ \ignorespaces
}{%
  \par\medskip
}

\title{Intrinsic Stochastic Successive Convexification on SE(3) for Chance Constrained 6-DOF Rendezvous}

\author{Fabio D'Onofrio\footnote{Corresponding author, fabiod@utexas.edu. Ph.D. candidate, Aerospace Engineering and Engineering Mechanics.} and Renato Zanetti\footnote{Associate Professor, Aerospace Engineering and Engineering Mechanics. AIAA Associate Fellow.}}
\affil{The University of Texas at Austin, Austin, Texas, 78712}

\begin{document}

\maketitle

\begin{abstract}
This work presents an intrinsic stochastic successive convexification method formulated on the Special Euclidean group SE(3) for six degrees of freedom spacecraft rendezvous trajectory optimization. The proposed approach extends stochastic successive convexification, originally developed for Euclidean state spaces, to the nonlinear manifold of SE(3), thereby enabling a consistent covariance steering and chance constrained optimization of rigid body pose trajectories. While conventional trajectory optimization methods often treat position and attitude separately, or account for stochastic dispersion only after a deterministic reference trajectory has been generated, the proposed SE(3)-based formulation captures the intrinsic coupling between translational and rotational motion uncertainty. This coupling is especially important for rendezvous problems with safety constraints that depend on the full relative pose, including collision avoidance, docking corridor, camera field of view, and probabilistic force and torque bounds. Numerical simulations show that jointly optimizing the nominal trajectory, covariance, and feedback law shapes the closed loop dispersion and improves probabilistic constraint satisfaction relative to tracking a deterministic reference with a feedback linearization controller.
\end{abstract}

\section*{Nomenclature}

{\renewcommand\arraystretch{1.0}
\noindent\begin{longtable*}{@{}l @{\quad=\quad} l@{}}
$6-DOF$ & six degrees of freedom \\
$DCM$ & direction cosine matrix \\
$ZOH$ & zero order hold \\
$FOV$ & field of view \\
$FOH$ & first order hold \\
$ICS$ & iterative covariance steering \\
$MLG$ & Matrix Lie Group \\
$MRP$ & modified Rodrigues parameters \\
$SOCP$ &stochastic optimal control problem \\
$SCP$ & sequential convex programming \\
$SCvx$ & successive convexification \\
$LMI$ & linear matrix inequality \\
$SDP$ & semidefinite program \\
SO(3) & special orthogonal group in three dimensions \\
SE(3) & special Euclidean group in three dimensions \\
\end{longtable*}}

\addtocounter{table}{-1}

\section{Introduction}
\lettrine{A}{utonomous} spacecraft operations, such as rendezvous, docking, proximity operations, and landing, require guidance algorithms that can rapidly generate translational and rotational maneuvers while satisfying safety critical constraints. Among the most challenging constraints from an algorithmic and implementation standpoint are those that inherently couple position and attitude, such as line of sight constraints between the controlled vehicle and a target. Whether the target is another spacecraft to be approached or a landing site, onboard sensors must remain properly pointed, while the thrust direction must remain within prescribed pointing constraints; otherwise, the vehicle may lose the information needed to update its trajectory in real time or violate sensor, actuator, or collision constraints \cite{RPOD_general,Descent_general}.

The Special Euclidean group SE(3) is the natural state space for rigid body pose: rather than treating position and attitude as unrelated variables, the Matrix Lie Group (MLG) SE(3) provides a fully coupled pose representation \cite{ChirBook2}. The same coupling carries over to guidance and control through the constraints and feedback laws that the optimizer must enforce. Field of view, line of sight, docking corridor, and sensor pointing requirements are functions of the relative pose, so their satisfaction depends jointly on translational and rotational states. This paper carries this pose level description into stochastic trajectory optimization: rigid body composition, relative pose errors, frame transformations, and coupled position attitude uncertainty and constraints are expressed directly on SE(3), while the control algorithm is built around perturbations of that pose.

In addition to satisfying geometric and operational constraints, the vehicle must operate reliably in the presence of navigation errors, actuation uncertainty, and unmodeled disturbances. Achieving further improvements in proximity operations and landing accuracy will therefore require guidance and control architectures that are robust to uncertainty, which can degrade estimation accuracy and tracking performance \cite{robustdesc}.

One approach is to combine deterministic trajectory generation with robust feedback control. In this architecture, a nominal trajectory is first generated using a simplified guidance model, and a feedback controller is then designed to track that trajectory despite modeling error, parameter uncertainty, sensor noise, and external disturbances. For powered descent, this separation has been used to combine convex propellant optimal guidance with multivariable robust control, thereby providing robustness at the tracking level rather than by explicitly optimizing the dispersion of the trajectory distribution \cite{robust}.

However, a deterministic trajectory that remains close to a constraint boundary may still be unacceptable in practice. Even with a stabilizing tracking controller, stochastic dispersion may cause the realized trajectory to violate collision avoidance, field of view, plume impingement, pointing, or actuator constraints. Robust trajectory optimization addresses this issue by assuming that disturbances lie in a bounded set and enforcing constraint satisfaction for the worst case realization \cite{robcntrbook}. This approach is attractive when reliable uncertainty bounds are available, but it can become overly conservative when the uncertainty set must cover navigation error, actuation error, and unmodeled dynamics simultaneously. Moreover, if uncertainty is considered only after a nominal trajectory has been generated, the planner cannot account for how the feedback law will shape the dispersion near safety critical boundaries. Stochastic trajectory optimization provides an alternative by modeling uncertainty probabilistically and replacing hard path/control constraints with chance constraints \cite{CS3}. In this setting, the optimizer plans not only a nominal trajectory, but also the evolution of the state distribution, typically through its first two moments, so that constraint violation probabilities remain below prescribed risk levels. Chance constrained sequential convex programming follows this idea by repeatedly linearizing nonlinear dynamics and constraints about a reference trajectory, propagating covariance through the resulting local model, and solving a convex subproblem in which probabilistic constraints are transcribed into deterministic convex constraints \cite{lew_CC_SCP_2020}.

Within this class of stochastic sequential convex programming methods, covariance steering incorporates feedback design directly into the optimization \cite{CS_1, CS_2}. Rather than treating covariance as the byproduct of a prescribed controller, covariance steering optimizes an affine feedback policy together with the nominal trajectory so that the first two moments of the stochastic system satisfy prescribed boundary and/or path requirements. For nonlinear dynamics and constraints, this gives rise to Iterative Covariance Steering (ICS), in which the stochastic optimal control problem is repeatedly approximated by a convex program in the mean, covariance, and feedback variables \cite{ICS}. This approach has been used for chance constrained stochastic trajectory optimization in Euclidean coordinates, including 6-DOF spacecraft rendezvous applications in which attitude is represented by modified Rodrigues parameters (MRP) \cite{zhang2023stoch6DOF}. Takubo and D'Amico extended this idea to stochastic attitude control on SO(3) through a mixed \textit{extrinsic/intrinsic} approach: the mean attitude evolves in the unit quaternion space (i.e. the extrinsic space), while the covariance is propagated in a local error modified Rodrigues parameters space (i.e. the intrinsic space). This avoids assigning covariance directly to the constrained quaternion coordinates or relying on directional statistics on SO(3), while enabling joint optimization of the mean attitude, local error covariance, and affine feedback gain within an iterative covariance steering framework \cite{takubo_damico_2024}.

These developments build on the broader literature on Sequential Convex Programming (SCP) for trajectory optimization. SCP denotes a broad class of methods that solve nonlinear optimal control problems through a sequence of convex approximations. Specific optimization algorithms in this class include SCvx and GuSTO \cite{malyuta_tutorial}. In SCvx, the nonconvex dynamics and constraints are linearized about a reference trajectory, \textit{virtual controls} (i.e. additional optimization variables, often called slack variables) are introduced to prevent \textit{artificial infeasibility} (i.e. when the convexified problem is unfeasible while the original nonconvex one was feasible), and \textit{trust region} and penalty mechanisms are used to promote convergence and avoid \textit{artificial unboundedness}, which is when linearization renders the solution unbounded from below \cite{SCvx}. For deterministic 6-DOF powered descent, this framework has been combined with state triggered constraints and with dual quaternions as a unified representation of position and attitude \cite{szmuk_6dof_2018,reynolds_DQ_2020}. Recent work on intrinsic SCP further emphasizes that, when the state evolves on a smooth manifold, the local correction solved by the convex subproblem should be expressed in the tangent space of the manifold rather than in redundant embedding coordinates \cite{intrinsicSCvx}. Thus, two ingredients have been developed in parallel: stochastic sequential convex programming methods that steer mean and covariance subject to chance constraints, and geometric sequential convex programming methods that account for the non Euclidean structure of rigid body pose.

The geometric machinery needed to combine these ingredients is summarized in Sec.~\ref{sec2}. There, the exponential and logarithm maps are used to move between a Matrix Lie Group and its Lie algebra, and the boxed plus (retraction) and boxed minus (inverse retraction) operators are introduced to apply and measure perturbations on the group. The nominal trajectory evolves on SE(3), while the mean corrections, covariance propagation, affine feedback law, and chance constraint transcription are all expressed in compatible tangent coordinates. This provides the control counterpart of the same geometric uncertainty modeling idea: the uncertainty model remains local and vector space valued, but the accepted pose trajectory remains on the manifold.

To the authors' knowledge, no existing work applies stochastic sequential convex programming, in the form of iterative covariance steering with chance constraints, to 6-DOF rigid body pose trajectory optimization or control using either unit dual quaternions or homogeneous transformation matrices on SE(3). The contribution of this work is therefore a geometrically consistent stochastic optimal control formulation in which the nominal pose update, uncertainty model, feedback policy, and probabilistic constraints are all expressed in the Lie algebra of the SE(3) pose manifold. 

At each iteration, pose perturbations are represented with the left invariant inverse retraction $\ominus_{G,l}$, the nonlinear dynamics are linearized in the resulting tangent coordinates, and the discretized model is embedded in a convex subproblem that optimizes the nominal trajectory, covariance sequence, and affine feedback gains. Gaussian chance constraints are transcribed using tangent space covariance projections, and the accepted mean update is mapped back to SE(3) through the left retraction $\oplus_{G,l}$.

Although this work ultimately implements the method using homogeneous matrices, the same derivation can be carried out with dual quaternions by using the corresponding dual quaternion retraction and inverse retraction. Both representations describe the same pose group and share the same 6 dimensional tangent space se(3). The homogeneous matrix formulation is therefore one computational realization of a stochastic sequential convex programming framework for pose trajectories on SE(3). 

The remainder of the paper is organized as follows. Sections \ref{sec2} and \ref{sec:prob_mlg} summarize the Matrix Lie Group and the \textit{concentrated probability distribution} used by the proposed formulation, respectively. Section \ref{sec:problem_statement} states the original nonlinear chance constrained rendezvous problem. Section \ref{sec:chap4_sec2} presents the proposed intrinsic stochastic SCP solution method. Section \ref{sec:results} presents simulation results and Section \ref{sec6} concludes the paper.

\section{Matrix Lie Groups}\label{sec2}

This section briefly introduces only the Matrix Lie Group concepts used by the proposed formulation. More complete treatments can be found in \cite{hall2003,ChirBook2,Sola}.

A real Matrix Lie Group $G$ is a subgroup of $GL_n(\mathbb{R})$ that is also an embedded smooth manifold. Its tangent space at the identity, denoted by $\mathcal{G}=T_I G$, is the Lie algebra of $G$ and is isomorphic to $\mathbb{R}^N$, where $N=\dim G$. The maps $[\cdot]_G^\wedge:\mathbb{R}^N\rightarrow\mathcal{G}$ and $[\cdot]_G^V:\mathcal{G}\rightarrow\mathbb{R}^N$ denote the ''wedge'' and ''vee'' maps between vector coordinates and algebra elements.

The Special Euclidean group SE(3) is:

\begin{equation}\label{eq:SE(3)}
SE(3)
=
\left\{
T\in\mathbb{R}^{4\times4}
\mid
T=
\begin{bmatrix}
R & \bm{t}\\
\bm{0}_{1\times3} & 1
\end{bmatrix},
\;R\in SO(3),
\;\bm{t}\in\mathbb{R}^3
\right\}
\end{equation}

where SO(3) is the MLG containing all rotation matrices, i.e. $SO(3)=\{R\in\mathbb{R}^{3\times3}\mid R^\top R=I,\det(R)=1\}$. 

The corresponding Lie algebra se(3) has dimension six, as the number of degrees of freedom for a rigid body in 3D space.

The matrix exponential $\exp_G: \mathcal{G} \rightarrow G$ and logarithm $\log_G: G \rightarrow \mathcal{G}$ maps define a local correspondence between the Lie algebra and the group. The capitalized exponential and logarithm maps used throughout the paper are defined as~\cite{Sola}:

\begin{equation}
    \mathrm{Exp}_G(\bm{\tau})
    =
    \exp_G\!\left([\bm{\tau}]_G^\wedge\right) : \mathbb{R}^N \rightarrow G,
    \qquad
    \mathrm{Log}_G(X)
    =
    \left[\log_G(X)\right]_G^V : G \rightarrow \mathbb{R}^N
    \label{eq:mlg_exp_log}
\end{equation}

where $\mathrm{Exp}_G(\cdot)$ takes a tangent vector $\bm\tau \in \mathbb{R}^N$ into the corresponding SE(3) homogeneous matrix, whereas $\mathrm{Log}_G(\cdot)$ takes an SE(3) matrix and outputs the corresponding Lie algebra tangent vector $\bm\tau$.

Given a reference element $X\in G$ and a vector perturbation $\bm{\tau}\in\mathbb{R}^N$, the left boxed plus (retraction) and boxed minus (inverse retraction) are:

\begin{equation}\label{eq:lboxplus}
Y = \bm{\tau}\oplus_{G,l} X =  \mathrm{Exp}_G(\bm{\tau}) X \in G
\end{equation}
\begin{equation}\label{eq:lboxminus}
\bm{\tau} = Y \ominus_{G,l} X = \mathrm{Log}_G(YX^{-1}) \in \mathbb{R}^N
\end{equation}

For SE(3), the maps used in the pose update are:

\begin{equation}\label{eq:SE3hat}
\bm{\xi} = \begin{bmatrix} \omega_x \\ \omega_y \\ \omega_z \\ v_x \\ v_y \\ v_z \end{bmatrix} = \begin{bmatrix} \bm{\omega} \\ \bm{v} \end{bmatrix} \in \mathbb{R}^6 \xrightarrow{\left[ \cdot \right]_{SE(3)}^{\wedge}} \mathcal{X} =  \left[ \bm{\xi} \right]_{SE(3)}^{\wedge} = \begin{bmatrix}
    0 & -\omega_z & \omega_y & v_x \\ \omega_z & 0 & -\omega_x & v_y \\ -\omega_y  &  \omega_x  &   0  & v_z \\ 0 & 0 & 0 & 0
\end{bmatrix} \in se(3)
\end{equation}

\begin{equation}\label{eq:SE3vee}
\mathcal{X} \in se(3) \xrightarrow{\left[ \cdot \right]_{SE(3)}^{V}} \bm{\xi} = \left[\mathcal{X}\right]_{SE(3)}^{V} = \begin{bmatrix} \omega_x \\ \omega_y \\ \omega_z \\ v_x \\ v_y \\ v_z \end{bmatrix} = \begin{bmatrix} \bm{\omega} \\ \bm{v} \end{bmatrix} \in \mathbb{R}^6
\end{equation}

\begin{equation}\label{eq:SE3exp}
\left[ \bm{\xi} \right]_{SE(3)}^{\wedge} \in se(3) \xrightarrow{\exp_{SE(3)}} T = \begin{bmatrix}
\text{exp}_{SO(3)}(\left[\bm{\omega}\right]^{\wedge}_{SO(3)}) &  J_{SO(3),l}(\bm{\omega}) \bm{v} \\ \bm{0}^T & 1
\end{bmatrix} \in SE(3)
\end{equation}

\begin{equation}\label{eq:SE3log}
T \in SE(3) \xrightarrow{\log_{SE(3)}} \left[ \bm{\xi} \right]_{SE(3)}^{\wedge} = \begin{bmatrix}
\text{log}_{SO(3)}(R) &  J_{SO(3),l}^{-1}(\text{Log}_{SO(3)}(R)) \bm{t} \\ \bm{0}^T & 0
\end{bmatrix} \in se(3)
\end{equation}

where $J_{SO(3),l}$ is the standard left Jacobian of SO(3), and its expression, along with the SO(3) exponential $\text{exp}_{SO(3)}$ and logarithm $\text{log}_{SO(3)}$ expressions, can be found in \cite{ChirBook2}. 

The full state used in the rendezvous problem combines the pose with Euclidean velocity variables. This is represented as the product group $SE(3)\times\mathbb{R}^{n_c}$: on the Euclidean component, the group operation is vector addition and the exponential, logarithm, wedge, and vee maps reduce to the identity.

\section{Probability Distributions on Matrix Lie Groups}\label{sec:prob_mlg}

Since MLGs are nonlinear spaces, Euclidean probability distributions cannot be assigned to group elements by simply treating their matrix entries as independent coordinates. 

Several alternatives are available. Directional statistics distributions, such as wrapped normal, von Mises, and Bingham distributions, account for the global structure of the nonlinear manifold \cite{dirStat}. Projected Gaussian constructions instead define local Gaussian coordinates and map them to the manifold by projection. A third approach, used in this work and referred to as  concentrated probability distribution, is to assume that the probability mass is concentrated in a neighborhood where the logarithm map is well defined and to represent uncertainty in the Lie algebra \cite{concPDF,barfoot}.

This concentrated distribution viewpoint is natural for trajectory optimization because the optimizer works with local perturbations around a nominal pose trajectory. The group element remains on $G$, while the uncertainty is represented by a vector in the tangent space. For unimodular MLGs, including SO($n$) and SE($n$) (i.e. the generic Special Orthogonal and Special Euclidean groups of dimension $n$), integration can be defined with respect to a bi-invariant Haar measure, which is locally expressed through exponential coordinates and the determinant of the group Jacobian \cite{applebaum}. Under the concentrated assumption, this Jacobian is close to the identity over the region carrying most of the probability mass, so the local distribution can be manipulated using the same covariance algebra used in Euclidean space.

This paper uses the left concentrated convention. For a probability density $p(X)$ on $G$, the left group theoretic mean $\mu_l$ is defined by:

\begin{equation}
    \int_{G'}
    \left(
        X \ominus_{G,l} \mu_l
    \right)
    p(X)dX
    =
    \int_{G'}
    \mathrm{Log}_{G}
    \left(
        X\mu_l^{-1}
    \right)
    p(X)dX
    =
    \bm{0}
    \label{eq:mlg_left_mean}
\end{equation}

where $G'\subseteq G$ is a neighborhood of $\mu_l$ which contains the support of $p$ and is such that $\mathrm{Log}_{G}(X\mu_l^{-1})$ is well defined for all $X\in G'$. This is a group theoretic mean, distinct in general from the Fr{\'e}chet mean obtained by minimizing squared geodesic distance \cite{MLGdefMean,ManifoldMeans}. The corresponding left covariance is:
\begin{equation}
    P
    =
    \int_{G'}
    \left[
        X \ominus_{G,l} \mu_l
    \right]
    \left[
        X \ominus_{G,l} \mu_l
    \right]^\top
    p(X)dX
    \in
    \mathbb{R}^{N\times N}
    \label{eq:chap4_tangent_covariance}
\end{equation}

If $\bm{\tau}\sim\mathcal{N}(\bm{0},P)$ in $\mathbb{R}^N$ and:

\begin{equation}
    X
    =
    \bm{\tau}\oplus_{G,l}\mu_l
    =
    \mathrm{Exp}_{G}(\bm{\tau})\mu_l
    \label{eq:mlg_left_cgd_sample}
\end{equation}

then $X$ is denoted $X\sim\mathcal{N}_l(\mu_l,P)$, a left concentrated Gaussian distribution on $G$. Its local density can be written as:

\begin{equation}
    p(X)
    =
    c
    \exp
    \left(
        -\frac{1}{2}
        \mathrm{Log}_{G}(X\mu_l^{-1})^\top
        P^{-1}
        \mathrm{Log}_{G}(X\mu_l^{-1})
    \right)
    \label{eq:mlg_left_cgd_pdf}
\end{equation}
with $c\approx\left((2\pi)^N\det P\right)^{-1/2}$ when the support is sufficiently concentrated.

\section{Problem Statement}\label{sec:problem_statement}

The rendezvous example considers a chaser spacecraft approaching a passive target spacecraft on a circular orbit. The scenario is the same as the one in \cite{zhang2023stoch6DOF}. A Hill frame $\{H\}$ is attached to the target, with the $x_H$ axis pointing radially outward, the $y_H$ axis aligned with the target orbital velocity, and the $z_H$ axis completing the right-handed triad. The chaser body frame is denoted by $\{C\}$. The relative pose of $\{C\}$ with respect to $\{H\}$ is represented by the homogeneous transformation:

\begin{equation}
    T_{CH}
    =
    \begin{bmatrix}
        R_{CH} & \bm{\rho}^H\\
        \bm{0}^\top & 1
    \end{bmatrix}
    \in \mathrm{SE}(3)
    \label{eq:chap4_ren_pose}
\end{equation}

where $\bm{\rho}^H \in \mathbb{R}^3$ is the relative position expressed in the Hill frame and $R_{CH} \in SO(3)$ is the rotation matrix from the chaser body frame to the Hill frame. The embedded state is:

\begin{equation}
    \bm{x}
    =
    \begin{bmatrix}
        \mathrm{vec}(T_{CH}) \\
        \bm{\omega}_{CI}^C \\
        \bm{v}^H
    \end{bmatrix}
    \in \mathbb{R}^{22}
    \label{eq:chap4_ren_state}
\end{equation}

where $\mathrm{vec}(T_{CH}) \in \mathbb{R}^{16}$, $\bm{\omega}_{CI}^C \in \mathbb{R}^3$ is the chaser angular velocity in the inertial frame $\{I\}$ and expressed in $\{C\}$ and $\bm{v}^H \in \mathbb{R}^3$ is the chaser velocity w.r.t. the target and expressed in $\{H\}$.

The translational motion follows the Hill--Clohessy--Wiltshire (HCW) equations \cite{HCW}:

\begin{equation}
    \dot{\bm{\rho}}^H
    =
    \bm{v}^H,
    \qquad
    \dot{\bm{v}}^H
    =
    A_p\bm{\rho}^H
    +
    A_v\bm{v}^H
    +
    \frac{\bm{F}^H}{m}
    \label{eq:chap4_ren_hcw}
\end{equation}

with:

\begin{equation}
    A_p
    =
    \begin{bmatrix}
        3n^2 & 0 & 0\\
        0 & 0 & 0\\
        0 & 0 & -n^2
    \end{bmatrix},
    \qquad
    A_v
    =
    \begin{bmatrix}
        0 & 2n & 0\\
        -2n & 0 & 0\\
        0 & 0 & 0
    \end{bmatrix}
    \label{eq:chap4_ren_hcw_mats}
\end{equation}

where $\bm{F}^H \in \mathbb{R}^3$ is the commanded thrust expressed in the Hill frame, $m$ is the chaser mass and $n$ is the target mean motion, so that the angular velocity of the Hill frame w.r.t. the inertial frame $\{I\}$ in $\{H\}$ coordinates is $\bm{\omega}_{HI}^H=[0,0,n]^\top$.

The relative angular velocity entering the pose kinematics can be obtained as:

\begin{equation}
    \bm{\omega}_{CH}^C
    =
    \bm{\omega}_{CI}^C
    -
    C_{CH}\bm{\omega}_{HI}^H 
    \label{eq:chap4_ren_rel_omega}
\end{equation}

where $C_{CH}=R_{CH}^T$.

The pose kinematics are therefore driven by $\bm{\omega}_{CH}^C$ and $\bm{v}^C = C_{CH}\bm{v}^H$:

\begin{equation}
    \dot{T}_{CH}
    =
    T_{CH}
    \begin{bmatrix}
        [\bm{\omega}_{CH}^C]^\times & \bm{v}^C\\
        \bm{0}^\top & 0
    \end{bmatrix}
    \label{eq:chap4_ren_pose_kinematics}
\end{equation}

while the rotational dynamics are:

\begin{equation}
    \dot{\bm{\omega}}_{CI}^C
    =
    J^{-1}
    \left(
        -[\bm{\omega}_{CI}^C]^\times J\bm{\omega}_{CI}^C
        +
        \bm{\tau}^C
    \right)
    \label{eq:chap4_ren_euler}
\end{equation}

where $J \in \mathbb{R}^{3 \times 3}$ is the chaser inertia matrix expressed in the chaser body frame and $\bm{\tau}^C \in \mathbb{R}^3$ is the chaser torque input.

The control vector is then:

\begin{equation}
    \bm{u}
    =
    \begin{bmatrix}
        \bm{\tau}^C \\
        \bm{F}^H 
    \end{bmatrix} \in \mathbb{R}^6
    \label{eq:chap4_ren_control}
\end{equation}

Although the general stochastic optimal control formulation allows objectives based on expected cost or cost quantiles \cite{KumagaiCR3BP}, the rendezvous problem considered here follows the common chance constrained trajectory optimization approach of minimizing nominal control effort subject to probability bounds on the path and control constraints. Thus, uncertainty enters the optimization through the covariance dynamics, feedback gains, and chance constraints, whereas the trajectory cost measures the effort associated with the mean or feedforward maneuver. This choice separates performance from risk: fuel like control usage is optimized on the nominal trajectory, while robustness is imposed through explicit probability bounds on constraint satisfaction.

The original trajectory cost minimized in the rendezvous problem is then the cumulative control magnitude. Since force and torque have different units and numerical ranges, the torque contribution is converted to a force equivalent scale using the characteristic length $\ell_u=\tau_{\max}/F_{\max}$:

\begin{equation}
    J_{\mathrm{traj}}
    =
    \int_0^1
    \left(
        \left\|\bm{F}^H(\tau)\right\|_2
        +
        \frac{1}{\ell_u}
        \left\|\bm{\tau}^C(\tau)\right\|_2
    \right)
    d\tau
    \label{eq:chap4_ren_original_cost}
\end{equation}

where normalized time $\tau \in \left[0,1\right]$ is used.

The rendezvous path constraints describe collision avoidance, docking corridor entry, and sensor pointing. They are written as nonlinear scalar inequalities $g_j(\bm{x})\leq0$ in the original stochastic problem. Let $C_{TH}$ denote the DCM from the Hill frame to the target frame, obtained from the known target attitude. The collision model uses the Hill frame relative position $\bm{\rho}^H$ and the target frame position $\bm{\rho}^T=C_{TH}\bm{\rho}^H$.

The collision model follows the state triggered construction used by Zhang et al.~\cite{zhang2023stoch6DOF}. The first collision zone is a spherical keep out region centered at the target. This constraint represents the coarse safety sphere around the target spacecraft. The second collision zone models the solar panel envelope with an ellipsoid in the target frame. These two keep out constraints are enforced only while the chaser is outside the docking corridor: once the chaser is inside the prescribed docking cone, the protected neighborhood can be entered for terminal approach. 

The spherical keep out zone inequality constraint is:

\begin{equation}
    g_{\mathrm{sph}}
    =
    r_{\mathrm{KOZ}}^2
    -
    \left(\bm{\rho}^H\right)^\top\bm{\rho}^H
    \leq
    0
    \label{eq:chap4_ren_sphere_constraint}
\end{equation}

The solar panel ellipsoid keep out zone inequality constraint is:

\begin{equation}
    g_{\mathrm{pan}}
    =
    1
    -
    \left(\bm{\rho}^T\right)^\top
    H_{\mathrm{pan}}
    \bm{\rho}^T
    \leq
    0,
    \qquad
    H_{\mathrm{pan}}
    \succ
    0
    \label{eq:chap4_ren_panel_constraint}
\end{equation}

where $H_{\text{pan}} = C_{TH}^\top \, \text{diag}(1/r_x^2, 1/r_y^2, 1/r_z^2) \, C_{TH}$, with $r_x, r_y, r_z$ the semi-major axes of the ellipsoid.

The docking corridor is the trigger condition. The cone has vertex $\bm{p}_{\mathrm{dock}}^T$, axis $\bm{e}_{\mathrm{dock}}^T$, and half angle $\beta_{\mathrm{dock}}$. Defining:

\begin{equation}
    \bm{r}_{\mathrm{dock}}^T
    =
    C_{TH}\bm{\rho}^H
    -
    \bm{p}_{\mathrm{dock}}^T
    \label{eq:chap4_ren_dock_relative_position}
\end{equation}

the docking corridor inequality constraint is:

\begin{equation}
    g_{\mathrm{dock}}
    =
    \cos\beta_{\mathrm{dock}}
    -
    \left(\bm{e}_{\mathrm{dock}}^T\right)^\top
    \frac{\bm{r}_{\mathrm{dock}}^T}
    {\left\|\bm{r}_{\mathrm{dock}}^T\right\|}
    \leq
    0
    \label{eq:chap4_ren_docking_constraint}
\end{equation}

Thus, $g_{\mathrm{dock}}>0$ means that the chaser is outside the docking cone, while $g_{\mathrm{dock}}\leq0$ means that it is inside the approach corridor. The phase-dependent collision logic is:

\begin{equation}
    g_{\mathrm{dock}}\geq0
    \Rightarrow
    g_{\mathrm{sph}}\leq0
    \quad
    \mathrm{and}
    \quad
    g_{\mathrm{pan}}\leq0
    \label{eq:chap4_ren_collision_logic}
\end{equation}

This implication is embedded into the continuous optimization problem through state triggered functions:

\begin{equation}
    h_{\mathrm{sph}}
    =
    -\min\{-g_{\mathrm{dock}},0\}
    g_{\mathrm{sph}},
    \qquad
    h_{\mathrm{pan}}
    =
    -\min\{-g_{\mathrm{dock}},0\}
    g_{\mathrm{pan}}
    \label{eq:chap4_ren_triggered_collision}
\end{equation}

When $g_{\mathrm{dock}}>0$, the multiplier is positive and $h_{\mathrm{sph}}\leq0$, $h_{\mathrm{pan}}\leq0$ reduce to the sphere and panel keep out constraints. When $g_{\mathrm{dock}}\leq0$, both triggered constraints are inactive, allowing the chaser to enter the protected region through the docking corridor. In the stochastic discrete transcription, the trigger is evaluated on the reference mean state. Thus, at node $k$, the chance constraint associated with each keep out function $g_j$, $j\in\{\mathrm{sph},\mathrm{pan}\}$, is implemented as:

\begin{equation}
    \tilde{h}_{j,k}
    =
    \begin{cases}
        \mathrm{Pr}\!\left[g_j(\bm{x}_k)\leq0\right]\geq1-\epsilon_{\mathrm{path}}
        &
        g_{\mathrm{dock}}(\bar{\bm{x}}_k)\geq0
        \\
        \mathrm{Pr}\!\left[g_{\mathrm{dock}}(\bm{x}_k)\leq0\right]\geq1-\epsilon_{\mathrm{path}}
        &
        g_{\mathrm{dock}}(\bar{\bm{x}}_k)<0
    \end{cases}
    \label{eq:chap4_ren_triggered_collision_piecewise}
\end{equation}

where $\epsilon_{\mathrm{path}}\in(0,1)$ is the allocated probability of violation for each active path constraint.

When the reference mean is outside the docking cone, the random trajectory is constrained to remain outside the sphere and panel keep out regions. Once the reference mean is inside the cone, the random trajectory is instead constrained to satisfy the docking cone inequality.

The camera field of view constraint is expressed in the chaser frame. The camera is located at $\bm{p}_{\mathrm{cam}}^C$ and points along $\bm{e}_{\mathrm{cam}}^C$. With:

\begin{equation}
    \bm{r}_{\mathrm{cam}}^C
    =
    -C_{CH}\bm{\rho}^H
    -
    \bm{p}_{\mathrm{cam}}^C
    \label{eq:chap4_ren_camera_los}
\end{equation}

the sensor pointing inequality constraint is then:

\begin{equation}
    g_{\mathrm{FOV}}
    =
    \cos\beta_{\mathrm{FOV}}
    -
    \left(\bm{e}_{\mathrm{cam}}^C\right)^\top
    \frac{\bm{r}_{\mathrm{cam}}^C}
    {\left\|\bm{r}_{\mathrm{cam}}^C\right\|}
    \leq
    0
    \label{eq:chap4_ren_fov_constraint}
\end{equation}
This constraint couples translation and attitude because the line of sight vector depends on both $\bm{\rho}^H$ and $C_{CH}$.

For the stochastic problem, the FOV constraint is then:

\begin{equation}
    \mathrm{Pr}
    \left[
        g_{\mathrm{FOV}}(\bm{x}_k)
        \leq
        0
    \right]
    \geq
    1-\epsilon_{\mathrm{path}}
    \label{eq:chap4_ren_path_cc_prob}
\end{equation}

The control constraints are also enforced as chance constraints. For each physical control block $\bm{u}_{c,k}\in\{\bm{\tau}_k^C,\bm{F}_k^H\}$ with limit $u_{c,\max}\in\{\tau_{\max},F_{\max}\}$, the desired probabilistic constraint is:

\begin{equation}
    \mathrm{Pr}
    \left[
        \left\|\bm{u}_{c,k}\right\|_2
        \leq
        u_{c,\max}
    \right]
    \geq
    1-\epsilon_{\mathrm{ctrl}}
    \label{eq:chap4_ren_control_cc_prob}
\end{equation}

where $\epsilon_{\mathrm{ctrl}}\in(0,1)$ is the allocated probability of violation for each control magnitude constraint.

The values $\epsilon_{\mathrm{path}}$ and $\epsilon_{\mathrm{ctrl}}$ are individual risk allocations. If a joint chance constraint with total risk $\epsilon$ is desired over a collection of scalar constraints, a conservative transcription can be obtained with Boole's inequality by assigning individual risks $\epsilon_i$ such that:

\begin{equation}
    \sum_{i=1}^{n_g}\epsilon_i
    \leq
    \epsilon
    \label{eq:ren_boole_risk_allocation}
\end{equation}

and enforcing each scalar chance constraint separately at level $1-\epsilon_i$. A common simple choice is uniform allocation, $\epsilon_i=\epsilon/n_g$, although optimized or problem dependent risk allocation can reduce conservatism \cite{optrisk}. In this work, the risk levels are prescribed rather than optimized. 

The numerical values used in the simulation are reported in Sec.~\ref{sec:results}.

The full nonlinear chance constrainted rendezvous stochastic optimal control problem is summarized below.

\begin{problem}[Original nonlinear chance constrained rendezvous SOCP]\label{prob:ren_original}
Find the nominal trajectory, feedforward control, covariance trajectory, and affine feedback policy that solve
\begin{equation}
\begin{aligned}
    \min_{\bm{x}(\cdot),\,\bm{u}(\cdot),\,P(\cdot),\,K(\cdot)}\quad&
    J_{\mathrm{traj}}
    \qquad
    \mbox{Eq. (\ref{eq:chap4_ren_original_cost})}
    \\
    \mathrm{subject\;to:}\quad&
    \text{nonlinear }\mathrm{SE}(3)\text{ rendezvous dynamics, }
    \mbox{Eqs. (\ref{eq:chap4_ren_hcw}), (\ref{eq:chap4_ren_pose_kinematics}), and (\ref{eq:chap4_ren_euler}) }
    \\
    &\bm{x}(0)=\bm{x}_0,\qquad \bm{x}(1)=\bm{x}_f,
    \\
    &P(0)=P_0,\qquad P(1)\preceq P_f,
    \\
        &\begin{cases}
    \mathrm{Pr}\!\left[g_{\mathrm{sph}}(\bm{x}(\tau))\leq0\right]\geq1-\epsilon_{\mathrm{path}},
    \quad
    \mathrm{Pr}\!\left[g_{\mathrm{pan}}(\bm{x}(\tau))\leq0\right]\geq1-\epsilon_{\mathrm{path}},
    &
    g_{\mathrm{dock}}(\bar{\bm{x}}(\tau))\geq0,
    \\
    \mathrm{Pr}\!\left[g_{\mathrm{dock}}(\bm{x}(\tau))\leq0\right]\geq1-\epsilon_{\mathrm{path}},
    &
    g_{\mathrm{dock}}(\bar{\bm{x}}(\tau))<0,
    \end{cases}
    \qquad \tau\in[0,1],
    \\
    &\mathrm{Pr}\!\left[g_{\mathrm{FOV}}(\bm{x}(\tau))\leq0\right]\geq1-\epsilon_{\mathrm{path}},
    \qquad \tau\in[0,1],
    \\
    &\mathrm{Pr}\!\left[\|\bm{\tau}^C(\tau)\|_2\leq\tau_{\max}\right]\geq1-\epsilon_{\mathrm{ctrl}},
    \qquad \tau\in[0,1],
    \\
    &\mathrm{Pr}\!\left[\|\bm{F}^H(\tau)\|_2\leq F_{\max}\right]\geq1-\epsilon_{\mathrm{ctrl}},
    \qquad \tau\in[0,1]
\end{aligned}
\label{eq:ren_original_ocp}
\end{equation}

where the collision chance constraints are enforced through the continuous-time state-triggered formulation above, whose discrete implementation is given in Eq. (\ref{eq:chap4_ren_triggered_collision_piecewise}); $g_{\mathrm{FOV}}$ is defined in Eq. (\ref{eq:chap4_ren_fov_constraint}), and $P_0$, $P_f$ are prescribed initial and terminal covariances.

\end{problem}

\section{Intrinsic stochastic successive convexification}\label{sec:chap4_sec2}

This section presents the proposed intrinsic stochastic successive convexification (isSCvx) algorithm. The method extends the SCvx algorithm to the stochastic case and to non Euclidean state spaces: the continuous-time problem is discretized, the nonlinear dynamics and constraints are approximated about a reference trajectory, and a sequence of convex subproblems is solved with trust region and virtual control safeguards \cite{SCvx1,malyuta_tutorial}. Chance constraints are handled through deterministic sufficient conditions involving means, covariances, and risk allocation, following standard chance constrained SCP and covariance steering constructions \cite{lew_CC_SCP_2020,ICS2}.

Let the state space be:

\begin{equation}
    G = SE(3)\times\mathbb{R}^{n_c}
    = \left\{ (T, \bm{x}_c) \mid T \in SE(3) \, , \, \bm{x}_c \in \mathbb{R}^{n_c} \right\}
\end{equation}

where the cartesian part of the state $\bm{x}_c$ collects the Euclidean states. For a reference state $\bar{\bm{x}}\in G$, the convex subproblem optimizes a tangent mean correction $\delta\bar{\bm{x}}_k\in\mathbb{R}^{n_t}$, with $n_t=6+n_c$, rather than the embedded state coordinates. The random tangent perturbation about this mean correction is denoted $\delta\bm{x}_k\in\mathbb{R}^{n_t}$.

For the rendezvous problem, local uncertainty and optimization corrections are represented in the 12-dimensional tangent state:

\begin{equation}
    \delta\bm{x}
    =
    \begin{bmatrix}
        \bm{\xi}_{\theta} \\
        \bm{\xi}_{\rho} \\
        \delta\bm{\omega} \\
        \delta\bm{v}
    \end{bmatrix} \in \mathbb{R}^{12}
    \label{eq:chap4_ren_tangent_state}
\end{equation}

where $\bm{\xi}_{\theta}$ and $\bm{\xi}_{\rho}$ are the attitude and translational components of the intrinsic pose perturbation induced by the left inverse retraction, while $\delta\bm{\omega}$ and $\delta\bm{v}$ are the chaser absolute angular velocity and relative velocity perturbations, respectively.

The manifold valued state is reconstructed through the left retraction:

\begin{equation}
    \bm{x}
    =
    \delta\bm{x} \oplus_{G,l} \bar{\bm{x}}
    = 
    \text{Exp}_G(\delta\bm{x}) \circ_G \bar{\bm{x}}
    =
    \begin{bmatrix} \text{vec}(\text{Exp}_{SE(3)}(\delta\bm{x}(1:6)) \, \bar{T}) \\ \bar{\bm{x}}_c + \delta\bm{x}(7:12) \end{bmatrix} \in \mathbb{R}^{22}
    \label{eq:chap4_geo_retraction}
\end{equation}

The corresponding inverse retraction is:

\begin{equation}
    \delta\bm{x}
    =
     \bm{x} \ominus_{G,l} \bar{\bm{x}}
     =
    \text{Log}_G(\bm{x} \circ_G \bar{\bm{x}}^{-1})
    =  \begin{bmatrix}  \mathrm{Log}_{SE(3)}(T \bar{T}^{-1}) \\ \bm{x}_c - \bar{\bm{x}}_c \end{bmatrix} \in \mathbb{R}^{12}
    \label{eq:chap4_geo_inverse_retraction}
\end{equation}

In the stochastic convex subproblem, the optimized tangent variable is the mean correction. The random tangent perturbation is decomposed as:

\begin{equation}
    \delta\bm{x}_k
    =
    \delta\bar{\bm{x}}_k
    +
    \tilde{\bm{x}}_k,
    \qquad
    \tilde{\bm{x}}_k
    \sim
    \mathcal{N}(\bm{0},P_k)
    \label{eq:chap4_tangent_perturbation_distribution}
\end{equation}

where $\delta\bar{\bm{x}}_k$ is the deterministic mean correction optimized by the convex subproblem and $\tilde{\bm{x}}_k$ is the zero-mean stochastic tangent perturbation. The corresponding candidate mean state is:

\begin{equation}
    \bar{\bm{x}}_k^{+}
    =
    \delta\bar{\bm{x}}_k
    \oplus_{G,l}
    \bar{\bm{x}}_k
    \label{eq:chap4_candidate_mean_state}
\end{equation}

and stochastic states are modeled with the left concentrated Gaussian distribution introduced in Sec.~\ref{sec:prob_mlg}:

\begin{equation}
    \bm{x}_k
    =
    \tilde{\bm{x}}_k
    \oplus_{G,l}
    \bar{\bm{x}}_k^{+},
    \qquad
    \bm{x}_k
    \sim
    \mathcal{N}_l(\bar{\bm{x}}_k^{+},P_k)
    \label{eq:chap4_left_concentrated_state}
\end{equation}

Equivalently, the zero-mean group perturbation satisfies $\mathrm{Exp}_G(\tilde{\bm{x}}_k)\sim\mathcal{N}_l(\mathbb{I},P_k)$. Thus, covariance matrices are defined in the same tangent coordinates as the optimized mean corrections.

\subsection{Algorithm overview}

At outer iteration $i$, the algorithm starts from a reference sequence over a mesh of $N-1$ intervals, $\{(\bar{\bm{x}}_k^{(i)},\bar{\bm{u}}_k^{(i)})\}_{k=1}^N$ and a trust region radius $\eta^{(i)}$ whose role and use will be explained in the later section. Time dilation is the positive scale factor that maps normalized time $\tau\in[0,1]$ to physical time $t$:

\begin{equation}
    dt
    =
    s(\tau)d\tau
    \qquad
    s(\tau)>0
    \label{eq:chap4_time_dilation}
\end{equation}

For fixed final time problems, $s(\tau) = t_f$ is prescribed. For free final time problems, $s(\tau)$ may be included as a decision variable so that the optimizer adjusts the physical duration of each normalized time interval. In this work, final time is fixed.

The algorithm proceeds through five steps:

\begin{enumerate}
\item First, the nonlinear tangent space dynamics linearization matrices are computed, and the manifold propagation defect is measured; 
\item Second, tangent, control, covariance, feedback, virtual control, and other slack variables are scaled for numerical conditioning;
\item Third, a convex subproblem is solved in the tangent mean corrections, feedforward controls, covariance matrices, feedback gains, virtual controls, and other slack variables; 
\item Fourth, the candidate mean trajectory is reconstructed on the manifold and the physical covariance and feedback gains are recovered;
\item Finally, the candidate is accepted or rejected according to the predicted improvement, actual improvement, and feasibility decrease. 
\end{enumerate}

Each one of these steps will be detailed in the following sections.

\subsection{Dynamics linearization and discretization}

The normalized time interval is discretized at nodes:

\begin{equation}
    \tau_k=\frac{k-1}{N-1},
    \qquad
    k=1,\ldots,N
    \label{eq:chap4_scp_nodes}
\end{equation}

The reference control is interpolated with first order hold (FOH):

\begin{equation}
    \bar{\bm{u}}(\tau)
    =
    \alpha(\tau)\bar{\bm{u}}_k
    +
    \bigl(1-\alpha(\tau)\bigr)\bar{\bm{u}}_{k+1},
    \qquad
    \alpha(\tau)
    =
    \frac{\tau_{k+1}-\tau}{\tau_{k+1}-\tau_k}
    \label{eq:chap4_scp_foh}
\end{equation}

although the same construction can be specialized to zero order hold (ZOH) by setting the second interpolation weight to zero.

The stochastic tangent perturbation dynamics are written as:

\begin{equation}
    d \delta\bm{x}
    =
    f_{\delta x}(\bm{x},\bm{u},\tau) d\tau
    +
    G_c d\bm{w}
    \label{eq:chap4_tangent_dynamics}
\end{equation}

where $f_{\delta x}: \mathbb{R}^{n_t} \times \mathbb{R}^{n_u} \times\mathbb{R}^{n} \rightarrow \mathbb{R}^{n_t}$ is the tangent drift, $\bm{x} \in G$ and the random tangent perturbation $\delta\bm{x} \in \mathbb{R}^{n_t}$ were defined as functions of the reference state $\bar{\bm{x}}$ in Eq. (\ref{eq:chap4_geo_retraction}) and Eq. (\ref{eq:chap4_geo_inverse_retraction}), respectively, $G_c \in \mathbb{R}^{n_t \times n_w}$ is the continuous-time diffusion matrix, and $\bm{w}$ is a standard unit strength $n_w$-dimensional Wiener process.

Linearizing the continuous-time tangent stochastic dynamics about the reference trajectory gives:

\begin{equation}
    d\delta\bm{x}
    =
    \left(
    A_{\delta x}(\tau)\delta\bm{x}
    +
    B_{\delta u}(\tau)\delta\bm{u}
    +
    \bm{c}_{\delta x}(\tau)
    \right)d\tau
    +
    G_c d\bm{w}
    \label{eq:chap4_tangent_linear_ct}
\end{equation}

with drift Jacobians:

\begin{equation}
    \begin{aligned}
    A_{\delta x}(\tau)
    &=
    \left.
    \frac{\partial}{\partial \bm{\eta}}
    f_{\delta x}\!\left(
        \bm{\eta}\oplus_{G,l}\bar{\bm{x}}(\tau),
        \bar{\bm{u}}(\tau),
        \tau
    \right)
    \right|_{\bm{\eta}=\bm{0}},
    \\
    B_{\delta u}(\tau)
    &=
    \left.
    \frac{\partial}{\partial \delta\bm{u}}
    f_{\delta x}\!\left(
        \bar{\bm{x}}(\tau),
        \bar{\bm{u}}(\tau)+\delta\bm{u},
        \tau
    \right)
    \right|_{\delta\bm{u}=\bm{0}}
    \end{aligned}
    \label{eq:chap4_tangent_jacobians}
\end{equation}

and affine residual:

\begin{equation}
    \bm{c}_{\delta x}(\tau)
    =
    f_{\delta x}\!\left(
        \bar{\bm{x}}(\tau),
        \bar{\bm{u}}(\tau),
        \tau
    \right)
    \label{eq:chap4_tangent_affine_residual}
\end{equation}

The state derivative is therefore taken with respect to a tangent perturbation applied through the left retraction.

Over each interval, the method integrates the state transition matrix $\Phi_{\delta x}$, the FOH input integrals, and the process noise contribution induced by the diffusion term. Let:

\begin{equation}
    Q_{\delta x}(\tau)
    =
    \begin{cases}
        s(\tau)G_cG_c^\top, & \text{if normalized time dilation is used},\\
        G_cG_c^\top, & \text{otherwise},
    \end{cases}
    \label{eq:chap4_tangent_noise_psd}
\end{equation}

The additive covariance accumulated over the interval is then given by:

\begin{equation}
    \Sigma_{w,k}
    =
    \int_{\tau_k}^{\tau_{k+1}}
    \Phi_{\delta x}(\tau_{k+1},\sigma)
    Q_{\delta x}(\sigma)
    \Phi_{\delta x}^\top(\tau_{k+1},\sigma)
    d\sigma 
    \label{eq:chap4_process_noise_gramian}
\end{equation}

Instead of forming this integral directly, the noise Gramian $W(\tau,\tau_k)$ is integrated together with the state transition matrix, so the propagated quantities satisfy:

\begin{align}
    \frac{d}{d\tau} \Phi_{\delta x}(\tau,\tau_k)
    &=
    A_{\delta x}(\tau)\Phi_{\delta x}(\tau,\tau_k) , \quad \Phi_{\delta x}(\tau_k,\tau_k) = I_{n_t \times n_t}
    \nonumber\\
    \frac{d}{d\tau} W(\tau,\tau_k)
    &=
    \Phi_{\delta x}(\tau,\tau_k)^{-1}Q_{\delta x}(\tau)\Phi_{\delta x}(\tau,\tau_k)^{-\top} , \quad W(\tau_k,\tau_k) = O_{n_t \times n_t}
    \label{eq:chap4_tangent_aug_ode}
\end{align}

The linearized discrete-time matrices are then:

\begin{align}
    A_k
    &=
    \Phi_{\delta x}(\tau_{k+1},\tau_k)
    \nonumber\\
    B_k^-
    &=
    A_k
    \int_{\tau_k}^{\tau_{k+1}}
    \Phi_{\delta x}^{-1}(\tau,\tau_k)
    \alpha(\tau)B_{\delta u}(\tau)\,d\tau
    \nonumber\\
    B_k^+
    &=
    A_k
    \int_{\tau_k}^{\tau_{k+1}}
    \Phi_{\delta x}^{-1}(\tau,\tau_k)
    \bigl(1-\alpha(\tau)\bigr)B_{\delta u}(\tau)\,d\tau
    \nonumber\\
    \Sigma_{w,k}
    &=  A_kW_kA_k^\top \doteq G_k G_k^\top
    \label{eq:chap4_tangent_disc}
\end{align}

where $W_k = W(\tau_{k+1},\tau_k)$ and $G_k \in \mathbb{R}^{n_t \times n_t}$ is the square root of $\Sigma_k$. 

The nonlinear propagation defect is measured intrinsically as:

\begin{equation}
    \bm{d}_k
    =
    \bm{x}_{k+1}^{\mathrm{prop}}
    \ominus_{G,l}
    \bar{\bm{x}}_{k+1}
    \label{eq:chap4_tangent_defect}
\end{equation}

where $\bm{x}_{k+1}^{\mathrm{prop}}$ is obtained by propagating the nonlinear dynamics from $\bar{\bm{x}}_k$ over $[\tau_k,\tau_{k+1}]$ under the interpolated reference control. The linearized tangent mean-correction dynamics are then written for $\delta\bar{\bm{x}}_k$:

\begin{equation}
    \delta\bar{\bm{x}}_{k+1}
    =
    A_k\delta\bar{\bm{x}}_k
    +
    B_k^-\delta\bm{u}_{\mathrm{ff},k}
    +
    B_k^+\delta\bm{u}_{\mathrm{ff},k+1}
    +
    \bm{d}_k
    \label{eq:chap4_tangent_mean_dyn}
\end{equation}
where $\delta\bm{u}_{\mathrm{ff},k}=\bm{u}_{\mathrm{ff},k}-\bar{\bm{u}}_k$ is the feedforward control perturbation.

In successive convexification, the linearized dynamics equality can make an intermediate convex subproblem infeasible when the reference trajectory is far from dynamically consistent or when the local approximation is too restrictive. A virtual control is therefore introduced as an artificial additive input in the convexified dynamics \cite{malyuta_tutorial}. This term is not a physical control, it is a slack variable for the dynamics defect. It is heavily penalized in the objective so that it restores feasibility during early iterations while being driven to zero as the reference trajectory becomes dynamically consistent. In the present formulation, the virtual control is defined in tangent coordinates as $\bm{\nu}_k\in\mathbb{R}^{n_t}$ and so the linearized tangent mean-correction dynamics constraints for each convex subproblem become:

\begin{equation}
    \delta\bar{\bm{x}}_{k+1}
    =
    A_k\delta\bar{\bm{x}}_k
    +
    B_k^-\delta\bm{u}_{\mathrm{ff},k}
    +
    B_k^+\delta\bm{u}_{\mathrm{ff},k+1}
    +
    \bm{d}_k
    +
    \bm{\nu}_k
    \label{eq:chap4_tangent_mean_dyn_subproblem}
\end{equation}

The feedback law used for covariance propagation and chance constraint evaluation must be written in the same tangent coordinates. Thus, instead of using the Euclidean difference $\bm{x}_k-\bar{\bm{x}}_k^{+}$, the zero-mean state error entering the controller is defined by the inverse retraction about the candidate mean. The applied control is:

\begin{equation}
    \bm{u}_k
    =
    \bm{u}_{\mathrm{ff},k}
    +
    K_k\tilde{\bm{x}}_k,
    \qquad
    \tilde{\bm{x}}_k
    =
    \bm{x}_k
    \ominus_{G,l}
    \bar{\bm{x}}_k^{+}
    \in
    \mathbb{R}^{n_t}
    \label{eq:chap4_tangent_feedback_policy}
\end{equation}

where $\bm{u}_{\mathrm{ff},k}$ is the feedforward control optimized by the convex subproblem and $K_k \in \mathbb{R}^{n_u \times n_t}$ is the optimal feedback gain mapping the zero-mean tangent state error to the control correction. This is the non Euclidean counterpart of the standard Euclidean affine feedback policy.

Next, the covariance recursion constraints are described.

Let the full discretized control matrix be $B_k = B_k^-+B_k^+$. With the tangent feedback law in Eq. (\ref{eq:chap4_tangent_feedback_policy}), the closed loop covariance recursion is:

\begin{equation}
    P_{k+1}
    =
    \left(A_k+B_kK_k\right)
    P_k
    \left(A_k+B_kK_k\right)^\top
    +
    \Sigma_{w,k}
    \label{eq:chap4_tangent_closed_loop_cov}
\end{equation}

Eq. (\ref{eq:chap4_tangent_closed_loop_cov}) is not convex in the decision variables $K_k$ and $P_k$ because of the products $K_kP_k$ and $K_kP_kK_k^\top$. Therefore, the auxiliary variable $Y_k=K_kP_k$ is introduced so that the cross term becomes affine in $(P_k,Y_k)$ \cite{Benedikter}. The remaining quadratic covariance update is then enforced through the Schur complement relaxation:

\begin{equation}
    \begin{bmatrix}
        P_{k+1}-\Sigma_{w,k}
        &
        F_k
        \\
        F_k^\top
        &
        P_k
    \end{bmatrix}
    \succeq
    0,
    \qquad
    F_k
    \doteq
    A_kP_k
    +
    B_kY_k 
    \label{eq:chap4_cov_lmi}
\end{equation}

For $P_k\succ0$, the Schur complement of the lower right block gives $P_{k+1}-\Sigma_{w,k}\succeq F_kP_k^{-1}F_k^\top$, which is the relaxed form of Eq. (\ref{eq:chap4_tangent_closed_loop_cov}) expressed in the tangent covariance coordinates used by the optimizer.

\subsection{Parameter update and scaling}\label{sec:chap4_parameter_scaling}

In order to avoid numerical problems for each convex subproblem solve, all variables are scaled/normalized. Let $S_t$ be the diagonal tangent state scaling matrix and let $S_u$ and $\bm{c}_u$ define the affine control scaling. The subproblem variables are:

\begin{equation}
    \delta\bar{\bm{x}}_k
    =
    S_t\delta\hat{\bar{\bm{x}}}_k,
    \qquad
    \bm{u}_{\mathrm{ff},k}
    =
    S_u\hat{\bm{u}}_k+\bm{c}_u
    \label{eq:chap4_mean_control_scaling}
\end{equation}

The reference control in scaled coordinates is:

\begin{equation}
    \bar{\hat{\bm{u}}}_k
    =
    S_u^{-1}\left(\bar{\bm{u}}_k-\bm{c}_u\right)
    \label{eq:chap4_reference_control_scaling}
\end{equation}

The dynamics constraint is also row scaled using the local magnitudes of the scaled dynamics matrices so that the equality residuals have comparable numerical weights.

Covariance variables are normalized using a square root of the initial covariance. Let $S_P$ satisfy $P_0=S_PS_P^\top$. The normalized covariance and covariance steering variables are:
\begin{equation}
    \hat{P}_k
    =
    S_P^{-1}P_kS_P^{-\top},
    \qquad
    \hat{Y}_k
    =
    Y_kS_P^{-\top}
    \label{eq:chap4_cov_normalization}
\end{equation}

With this normalization, $\hat{P}_0=I_{n_t \times n_t}$.

\subsection{Solve step}

At each outer iteration, the convex subproblem optimizes the scaled tangent mean corrections $\delta\hat{\bar{\bm{x}}}_k$, scaled feedforward controls $\hat{\bm{u}}_k$, scaled virtual controls $\hat{\bm{\nu}}_k$, normalized covariances $\hat{P}_k$, normalized feedback variables $\hat{Y}_k$, and any problem dependent additional slack variables. The candidate mean state is reconstructed after the solve as:

\begin{equation}
    \bar{\bm{x}}_k^{+}
    =
    \delta\bar{\bm{x}}_k^\star
    \oplus_{G,l}
    \bar{\bm{x}}_k
    \label{eq:chap4_candidate_reconstruction}
\end{equation}

and the physical covariance and feedback variables are recovered by:

\begin{equation}
    P_k^\star
    =
    S_P\hat{P}_k^\star S_P^\top,
    \qquad
    Y_k^\star
    =
    \hat{Y}_k^\star S_P^\top,
    \qquad
    K_k^\star
    =
    Y_k^\star\left(P_k^\star\right)^{-1}
    \label{eq:chap4_feedback_recovery}
\end{equation}

In scaled notation, the tangent dynamics constraint is imposed as:

\begin{equation}
    \delta\hat{\bar{\bm{x}}}_{k+1}
    =
    \hat{A}_k\delta\hat{\bar{\bm{x}}}_k
    +
    \hat{B}_k^-\left(\hat{\bm{u}}_k-\bar{\hat{\bm{u}}}_k\right)
    +
    \hat{B}_k^+\left(\hat{\bm{u}}_{k+1}-\bar{\hat{\bm{u}}}_{k+1}\right)
   +
    \hat{\bm{d}}_k
    +
    \hat{\bm{\nu}}_k
    \label{eq:chap4_scaled_tangent_mean_dyn}
\end{equation}

where $\hat{A}_k=S_t^{-1}A_kS_t$, $\hat{B}_k^\pm=S_t^{-1}B_k^\pm S_u$, $\hat{\bm{d}}_k=S_t^{-1}\bm{d}_k$, and $\hat{\bm{\nu}}_k=S_t^{-1}\bm{\nu}_k$. 

Because the dynamics, covariance recursion, and chance constraints are all local approximations about the current reference trajectory, the convex subproblem is not allowed to choose an arbitrarily large correction in one iteration. The trust region limits the scaled tangent state correction and, when enabled, the scaled change in feedforward control. This keeps the candidate trajectory inside the neighborhood where the linearized model is expected to be accurate and prevents artificial unboundedness of the convexified subproblem. At node $k$, this restriction is imposed as:

\begin{equation}
    \left\|\delta\hat{\bar{\bm{x}}}_k\right\|_{q_\mathrm{tr}}
    +
    \left\|\hat{\bm{u}}_k-\bar{\hat{\bm{u}}}_k\right\|_{q_\mathrm{tr}}
    \leq
    \eta
    \label{eq:chap4_tangent_trust_region}
\end{equation}

where $\eta>0$ is the current trust region radius and $q_\mathrm{tr}$ selects the norm used to measure the step, typically $1$, $2$, or $\infty$. The first term bounds the manifold update in scaled tangent coordinates, while the second term bounds the deviation of the feedforward control from the reference control. The control term can be omitted when only the state update is trusted. The radius $\eta$ is updated by the outer loop according to the agreement between the predicted and actual merit improvement, as it will be described in Sec.~\ref{sec:chap4_scp_outer_loop}.

The objective minimized by the convex solver has the generic form:

\begin{equation}
    J_\mathrm{sub}
    =
    \frac{1}{c_J}J_\mathrm{traj}^\mathrm{cvx}
    +
    J_\mathrm{vc}
    +
    J_\mathrm{slack}
    \label{eq:chap4_scp_subproblem_cost}
\end{equation}

Here $J_\mathrm{traj}^\mathrm{cvx}$ is the problem dependent convex trajectory cost, $c_J\geq1$ is a cost normalization, $J_\mathrm{vc}$ penalizes the tangent virtual control, and $J_\mathrm{slack}$ penalizes relaxation variables $\bm{s}_k$ introduced by any nonconvex constraint that required linearization. A representative virtual control penalty is:

\begin{equation}
    J_\mathrm{vc}
    =
    w_\mathrm{vc}
    \sum_{k=1}^{N-1}
    \left\|
        \hat{\bm{\nu}}_k
    \right\|_1
    \label{eq:chap4_scp_vc_cost}
\end{equation}

The boundary conditions, path constraints, control constraints, and cost are kept problem dependent. In the rendezvous example, these generic functions are specialized to the corresponding safety, pointing and actuator requirements.

Optional covariance size constraints can also be imposed along the trajectory, either as full matrix inequalities $\hat{P}_k\preceq\hat{P}_{\max}$ or as componentwise variance bounds.

\subsection{Chance constraints}
\label{sec:chap4_chance_constraints}

Consider an original nonlinear scalar chance constraint at node $k$:

\begin{equation}
    \mathrm{Pr}\!\left[
        g_j(\bm{x}_k)
        \leq
        0
    \right]
    \geq
    1-\epsilon_j
    \label{eq:chap4_state_chance_original}
\end{equation}

This constraint cannot be inserted directly into the convex subproblem, since the random variable $g_j(\bm{x}_k)$ is generally a nonlinear function of the stochastic trajectory. The first step is therefore to replace the nonlinear constraint by its local tangent approximation about the reference state. For a scalar path constraint $g_j(\bm{x}_k)\leq0$, the tangent linearization gives:

\begin{equation}
    g_j(\bm{x}_k)
    \approx
    g_j(\bar{\bm{x}}_k)
    +
    G_{j,k}\delta\bm{x}_k
    \label{eq:chap4_state_chance_lin}
\end{equation}

where $G_{j,k}\in\mathbb{R}^{1\times n_t}$ is the tangent Jacobian:

\begin{equation}
    G_{j,k}
    =
    \left.
    \frac{\partial}{\partial \delta\bm{x}}
    g_j\!\left(
        \delta\bm{x}
        \oplus_{G,l}
        \bar{\bm{x}}_k
    \right)
    \right|_{\delta\bm{x}=\bm{0}}
    \label{eq:chap4_state_chance_tangent_jacobian}
\end{equation}

The random tangent perturbation is represented by its mean correction and covariance. Thus, under the Gaussian tangent approximation, the linearized scalar constraint is a Gaussian random variable with mean:

\begin{equation}
    \mu_{g,j,k}
    =
    g_j(\bar{\bm{x}}_k)
    +
    G_{j,k}\delta\bar{\bm{x}}_k
    \label{eq:chap4_state_chance_projected_mean}
\end{equation}

and variance:

\begin{equation}
    \sigma_{g,j,k}^2
    =
    G_{j,k}P_kG_{j,k}^\top
    =
    G_{j,k}S_P\hat{P}_kS_P^\top G_{j,k}^\top
    \label{eq:chap4_state_chance_projected_variance}
\end{equation}

For a scalar Gaussian variable, the probability statement in Eq. (\ref{eq:chap4_state_chance_original}) is equivalent to:

\begin{equation}
    g_j(\bar{\bm{x}}_k)
    +
    G_{j,k}\delta\bar{\bm{x}}_k
    +
    \Phi_\mathrm{N}^{-1}(1-\epsilon_j)
    \sqrt{
        G_{j,k}S_P\hat{P}_kS_P^\top G_{j,k}^\top
    }
    \leq
    0
    \label{eq:chap4_state_chance_deterministic}
\end{equation}

where $\Phi_\mathrm{N}^{-1}$ is the inverse cumulative distribution function of the standard normal distribution. Eq. (\ref{eq:chap4_state_chance_deterministic}) is deterministic, but it is not yet in the form used by the convex subproblem because of the square root covariance margin. The implementation introduces an auxiliary slack variable $\xi_{j,k}$ so that the covariance projection is bounded by $\xi_{j,k}^2$, and it allows a nonnegative slack $s_{j,k}$ to avoid artificial infeasibility. The resulting deterministic convex constraints are:

\begin{align}
    G_{j,k}S_P\hat{P}_kS_P^\top G_{j,k}^\top
    &\leq
    2\bar{\xi}_{j,k}\xi_{j,k}
    -
    \bar{\xi}_{j,k}^2
    +
    \chi_{j,k}
    \nonumber\\
    g_j(\bar{\bm{x}}_k)
    +
    G_{j,k}\delta\bar{\bm{x}}_k
    +
    \Phi_\mathrm{N}^{-1}(1-\epsilon_j)\xi_{j,k}
    &\leq
    s_{j,k}
    \label{eq:chap4_state_chance_soc}
\end{align}

where $\bar{\xi}_{j,k}$ is the value from the previous accepted iterate. The first inequality in Eq. (\ref{eq:chap4_state_chance_soc}) replaces the nonconvex bound $G_{j,k}S_P\hat{P}_kS_P^\top G_{j,k}^\top\leq\xi_{j,k}^2$ with the affine tangent approximation of $\xi_{j,k}^2$ about $\bar{\xi}_{j,k}$, plus the slack $\chi_{j,k}\geq0$. Since the tangent line is a global underestimator of the convex function $\xi_{j,k}^2$, this replacement is conservative when $\chi_{j,k}=0$. The second inequality in Eq. (\ref{eq:chap4_state_chance_soc}) is the tightened nominal constraint. Inactive path constraints are assigned zero chance constraint relaxation at the corresponding node.

Control chance constraints follow the same idea, but the random quantity is now the norm of the closed loop control produced by the affine feedback law. For a control block $\bm{u}_{c,k}$ with maximum allowable magnitude $u_{\max}$, the original probabilistic requirement is $\mathrm{Pr}[\|\bm{u}_{c,k}\|_2\leq u_{\max}]\geq1-\epsilon$. Because $\bm{u}_{c,k}$ depends on the zero-mean tangent perturbation through feedback, its covariance depends on both the feedback gain and the state covariance. With the covariance steering change of variables, this feedback-induced stochastic spread is bounded by an auxiliary slack variable $\zeta_{c,k}$ through:

\begin{equation}
    \begin{bmatrix}
        \zeta_{c,k}I & \hat{Y}_{c,k}\\
        \hat{Y}_{c,k}^\top & \hat{P}_k
    \end{bmatrix}
    \succeq
    0
    \label{eq:chap4_control_variance_lmi}
\end{equation}

where $\hat{Y}_{c,k}$ contains the rows of $\hat{Y}_k$ associated with the control block. This LMI implies an upper bound on the largest eigenvalue of the control covariance, so the stochastic control magnitude can be conservatively limited by tightening the deterministic norm with a risk-dependent margin. The square root term in that margin is then replaced by its affine approximation about the previous iterate:

\begin{equation}
    \|\bm{u}_{\mathrm{ff},c,k}\|_2
    +
    \psi_\epsilon
    \left(
        \sqrt{\bar{\zeta}_{c,k}}
        +
        \frac{\zeta_{c,k}-\bar{\zeta}_{c,k}}
        {2\sqrt{\bar{\zeta}_{c,k}}}
    \right)
    \leq
    u_{\max}
    +
    s_{u,c,k}
    \label{eq:chap4_control_chance}
\end{equation}
where $\bm{u}_{\mathrm{ff},c,k}$ is the feedforward component of the selected control block, $\bar{\zeta}_{c,k}$ is taken from the previous iterate, $s_{u,c,k}\geq0$ is a relaxation variable, and $\psi_\epsilon$ is a risk dependent concentration factor. Thus, the deterministic nominal magnitude is tightened by a feedback-induced uncertainty margin before being compared directly with the physical control limit. For the three dimensional force and torque blocks used in the numerical study, $\psi_\epsilon=\sqrt{8\log(1/\epsilon)+3}$.

\subsection{Convex subproblem statement}

Collect all decision variables in:

\begin{equation}
    \mathcal{Z}
    =
    \left\{
    \delta\hat{\bar{\bm{x}}}_k,
    \hat{\bm{u}}_k,
    \hat{\bm{\nu}}_k,
    \hat{P}_k,
    \hat{Y}_k,
    \bm{s}_k
    \right\}_{k=1}^N
    \label{eq:chap4_decision_set}
\end{equation}

Here, $\bm{s}_k$ collects the auxiliary variables introduced to obtain convex deterministic transcriptions of nonconvex constraints, including epigraph variables such as $\xi_{j,k}$, covariance margin variables such as $\zeta_{c,k}$, and nonnegative convexification or constraint relaxation slacks such as $\chi_{j,k}$ and $s_{j,k}$.

For the rendezvous problem, the convex subproblem is obtained by inserting the problem specific cost, tangent dynamics, boundary conditions, and active path and control chance constraints into the decision set above. In the discretized convex subproblem, the trajectory cost integral is evaluated with trapezoidal quadrature:

\begin{equation}
    J_{\mathrm{traj}}^{\mathrm{cvx}}
    =
    \sum_{k=1}^{N}
    w_k
    \left(
        \left\|\bar{\bm{F}}_k^H\right\|_2
        +
        \frac{1}{\ell_u}
        \left\|\bar{\bm{\tau}}_k^C\right\|_2
    \right),
    \qquad
    w_1=w_N=\frac{1}{2(N-1)},
    \quad
    w_k=\frac{1}{N-1}
    \label{eq:chap4_ren_discrete_cost}
\end{equation}

where $\bar{\bm{F}}_k^H$ and $\bar{\bm{\tau}}_k^C$ denote the mean, or feedforward, control values optimized by the convex subproblem. The physical trajectory cost in Eq. (\ref{eq:chap4_ren_discrete_cost}) penalizes only the mean control effort. In the stochastic implementation, the augmented convex objective also includes the trapezoidal sum of the control spread epigraph variables $\zeta_{c,k}$ used in the control chance constraint transcription, together with the virtual control and constraint relaxation penalties. These additional terms regularize control dispersion and support convergence of the SCvx iteration.

For implementation of the tangent space dynamics constraints, the continuous-time drift used by the intrinsic discretization in normalized time is:

\begin{equation}
    \bm{f}_{\delta x}
    =
    t_f
    \begin{bmatrix}
        \bm{\omega}_{CH}^C\\
        C_{CH}\bm{v}^H\\
        J^{-1}\left(-[\bm{\omega}_{CI}^C]^\times J\bm{\omega}_{CI}^C+\bm{\tau}^C\right)\\
        A_p\bm{\rho}^H+A_v\bm{v}^H+\bm{F}^H/m
    \end{bmatrix}
    \label{eq:chap4_ren_tangent_drift}
\end{equation}

Linearizing Eq. (\ref{eq:chap4_ren_tangent_drift}) in the tangent state ordering of Eq. (\ref{eq:chap4_ren_tangent_state}) gives:

\begin{equation}
    \dot{\delta\bm{x}}
    =
    A_{\delta x}\delta\bm{x}
    +
    B_{\delta u}\delta\bm{u}
    +
    \bm{c}_{\delta x},
    \qquad
    \delta\bm{u}
    =
    \begin{bmatrix}
        \left(\delta\bm{\tau}^C\right)^\top &
        \left(\delta\bm{F}^H\right)^\top
    \end{bmatrix}^\top 
    \label{eq:chap4_ren_tangent_linearization}
\end{equation}

The affine residual is:

\begin{equation}
    \bm{c}_{\delta x}
    =
    \bm{f}_{\delta x}
    \left(
        \bar{\bm{x}},
        \bar{\bm{u}},
        \tau
    \right)
    \label{eq:chap4_ren_affine_residual}
\end{equation}

The Jacobians are:

\begin{equation}
    A_{\delta x}
    =
    t_f
    \begin{bmatrix}
        -[\bm{\omega}_{CI}^C]^\times & 0_{3\times3} & I_3 & 0_{3\times3}\\
        0_{3\times3} & -[\bm{\omega}_{CH}^C]^\times & 0_{3\times3} & C_{CH}\\
        0_{3\times3} & 0_{3\times3} & J^{-1}\left([J\bm{\omega}_{CI}^C]^\times-[\bm{\omega}_{CI}^C]^\times J\right) & 0_{3\times3}\\
        0_{3\times3} & A_p C_{CH}^\top & 0_{3\times3} & A_v
    \end{bmatrix}
    \label{eq:chap4_ren_A_tan}
\end{equation}

\begin{equation}
    B_{\delta u}
    =
    t_f
    \begin{bmatrix}
        0_{3\times3} & 0_{3\times3}\\
        0_{3\times3} & 0_{3\times3}\\
        J^{-1} & 0_{3\times3}\\
        0_{3\times3} & \frac{1}{m}I_3
    \end{bmatrix}
    \label{eq:chap4_ren_B_tan}
\end{equation}

Combining these components, the rendezvous convex subproblem solved at each SCP iteration is:

\begin{problem}[Rendezvous convex subproblem]\label{prob:ren_subproblem}
Find the nominal trajectory, feedforward control, covariance trajectory, and affine feedback policy that solve
\begin{equation}
\begin{aligned}
    \min_{\mathcal{Z}_{\mathrm{ren}}}
    \quad&
    \frac{1}{c_J}
    J_{\mathrm{traj}}^{\mathrm{cvx}}
    +
    w_{\mathrm{vc}}
    \sum_{k=1}^{N-1}
    \left\|
        \hat{\bm{\nu}}_k
    \right\|_1
    +
    J_{\mathrm{ren,slack}}
    \\[1mm]
    \mathrm{s.t.}\quad&
    \mbox{tangent mean-correction dynamics Eq. (\ref{eq:chap4_scaled_tangent_mean_dyn})}
    \quad
    k=1,\ldots,N-1
    \\
    &
    \mbox{mean and covariance boundary constraints}
    \\
    &
    \mbox{closed loop covariance recursion Eq. (\ref{eq:chap4_cov_lmi})}
    \quad
    k=1,\ldots,N-1
    \\
    &
    \mbox{active collision and FOV chance constraints Eq. (\ref{eq:chap4_state_chance_soc})}
    \\
    &
    \mbox{force and torque chance constraints Eq. (\ref{eq:chap4_control_variance_lmi}) and Eq. (\ref{eq:chap4_control_chance})}
    \\
    &
    \mbox{trust region constraints Eq. (\ref{eq:chap4_tangent_trust_region})}
\end{aligned}
\label{eq:chap4_ren_convex_subproblem}
\end{equation}
\end{problem}

Here, $\mathcal{Z}_{\mathrm{ren}}$ is the decision set in Eq. (\ref{eq:chap4_decision_set}) specialized to the rendezvous state and control dimensions. The active collision constraints are selected using the docking cone trigger described in Eq. (\ref{eq:chap4_ren_triggered_collision_piecewise}), and the FOV constraint is always included. The path and control chance constraints use the deterministic convex transcriptions derived in Sec.~\ref{sec:chap4_chance_constraints}. The rendezvous slack penalty is:

\begin{equation}
    J_{\mathrm{ren,slack}}
    =
    w_{\mathrm{nc}}
    \sum_{k=1}^{N}
    \left(
        \sum_{j\in\mathcal{A}_k}
        \left(
            s_{j,k}
            +
            \chi_{j,k}
        \right)
        +
        \sum_{c\in\{\tau,F\}}
        \chi_{c,k}
    \right)
    \label{eq:chap4_ren_slack_penalty}
\end{equation}

where $\mathcal{A}_k$ is the set of active path chance constraints at node $k$. The control spread epigraph variables $\zeta_{c,k}$ enter the augmented convex objective through the control chance constraint transcription as described in Sec.~\ref{sec:chap4_chance_constraints}.

Each convex subproblem is a semi-definite program (SDP) which can be efficiently solved with the interior point method.

\subsection{Outer loop update and convergence}
\label{sec:chap4_scp_outer_loop}

After each solve step, the candidate trajectory is repropagated through the nonlinear dynamics and assigned an augmented merit value:

\begin{equation}
    J_\mathrm{aug}
    =
    \frac{J_{\text{traj}}^{\text{cvx}}}{c_J}
    +
    J_\mathrm{dyn}
    +
    J_\mathrm{nc}
    \label{eq:chap4_stoch_merit}
\end{equation}

Here $J_{\text{traj}}^{\text{cvx}}$ is the original problem dependent trajectory cost evaluated without relaxation penalties, $J_\mathrm{dyn}$ penalizes the nonlinear tangent defects, and $J_\mathrm{nc}$ penalizes nonconvex or probabilistic constraint violations evaluated on the candidate solution with weight $w_{\mathrm{nc}}$. The dynamics penalty is:

\begin{equation}
    J_\mathrm{dyn}
    =
    w_{vc}
    \sum_{k=1}^{N-1}
    \left\|
        S_t^{-1}\bm{d}_k
    \right\|_1
    \label{eq:chap4_geo_dyn_pen}
\end{equation}

where $w_{vc} > 0$ is the same weight used to penalize virtual control in the convex subproblem (Eq. (\ref{eq:chap4_scp_vc_cost})) and $\bm{d}_k$ was defined in Eq. (\ref{eq:chap4_tangent_defect}).

To evaluate the feasibility of the candidate solution, four residuals are used. The dynamics residual is:

\begin{equation}
    \chi_\mathrm{dyn}
    =
    \max_{k=1,\ldots,N-1}
    \left\|
        S_t^{-1}\bm{d}_k
    \right\|_\infty 
    \label{eq:chap4_dyn_feas}
\end{equation}

The constraint residual $\chi_\mathrm{cc}$ is the largest normalized positive violation among the deterministic path, boundary, control, and chance constraints evaluated on the candidate trajectory. Thus, constraints that are satisfied contribute zero, while violated constraints contribute their positive residual after the same scaling used in the convex subproblem. The pathwise covariance residual $\chi_\mathrm{cov}$ measures violation of any covariance size constraints imposed along the trajectory, such as $P_k\preceq P_{\max}$ or componentwise variance bounds. 

The terminal covariance residual $\chi_{\mathrm{cov},f}$ similarly measures violation of the terminal covariance requirement $P_N\preceq P_f$.

These quantities are combined into a single feasibility metric:

\begin{equation}
    \chi
    =
    \max\left\{
        \chi_\mathrm{dyn},
        \chi_\mathrm{cc},
        \chi_\mathrm{cov},
        \chi_{\mathrm{cov},f}
    \right\}
    \label{eq:chap4_stoch_feas}
\end{equation}

A candidate is considered feasible by the algorithm when:

\begin{equation}
    \chi
    <
    \epsilon_\mathrm{feas}
    \label{eq:chap4_feasible_candidate}
\end{equation}

where $\epsilon_\mathrm{feas}$ is the feasibility tolerance. This test is intentionally based on the maximum residual: the candidate is feasible only when the dynamics defect, deterministic and chance constraint residuals, pathwise covariance residuals, and terminal covariance residual are all below tolerance. 

The predicted and actual merit improvements are then defined as:

\begin{equation}
    \Delta J_\mathrm{pred}
    =
    J_\mathrm{aug}
    -
    J_\mathrm{sub}^\star,
    \qquad
    \Delta J_\mathrm{act}
    =
    J_\mathrm{aug}
    -
    J_\mathrm{aug}^\star
    \label{eq:chap4_scp_improvements}
\end{equation}

The trust region ratio is introduced:

\begin{equation}
    \rho
    =
    \frac{\Delta J_\mathrm{act}}{\Delta J_\mathrm{pred}}
    \label{eq:chap4_scp_rho}
\end{equation}

The candidate is accepted and the trust region radius $\eta$ is grown of a factor $\beta_{grow} > 1$ if $\rho\geq\rho_2$, accepted and kept if $\rho_1\leq\rho<\rho_2$, accepted and shrunk with a factor $\beta_{shrink} < 1$ if $\rho_0\leq\rho<\rho_1$, and rejected otherwise. 

Convergence is declared only after an accepted step. The candidate must satisfy $\chi<\epsilon_\mathrm{feas}$ and at least one of the following two conditions must hold. The first is a small relative predicted improvement:

\begin{equation}
    \frac{\Delta J_\mathrm{pred}}
    {\left|J_\mathrm{aug}\right|}
    \leq
    \epsilon_\mathrm{rel}
    \label{eq:chap4_scp_rel_improv}
\end{equation}

The second is a small accepted step:

\begin{equation}
    \delta
    =
    \max\left\{
        \max_k
        \left\|
            S_t^{-1}
            \left(
                \bar{\bm{x}}_k^{+}\ominus_{G,l}\bar{\bm{x}}_k
            \right)
        \right\|_{q_\mathrm{exit}},
        \max_k
        \left\|
            \hat{\bm{u}}_k^\star-\bar{\hat{\bm{u}}}_k
        \right\|_{q_\mathrm{exit}}
    \right\}
    \leq
    \epsilon_\mathrm{abs}
    \label{eq:chap4_scp_step}
\end{equation}

where $q_\mathrm{exit}$ is the norm used to measure the accepted step for convergence.

The algorithm terminates without a convergence claim if the maximum iteration count is reached, or the trust region radius reaches its minimum value before a feasible accepted reference is obtained. The final output is a nominal trajectory on the manifold, a tangent covariance sequence, and an affine feedback policy. 

The complete algorithm is given in Algorithm \ref{AlgChap4GeometricSSCVX}.

\begin{algorithm}\caption{Intrinsic stochastic SCvx on $G=SE(3)\times\mathbb{R}^{n_c}$}\label{AlgChap4GeometricSSCVX}
\begin{algorithmic}[1]
\Input $\{\bar{\bm{x}}_k,\bar{\bm{u}}_k\}_{k=1}^N$, $\bm{\mu}_0$, $\bm{\mu}_f$, $P_0$, $P_f$, $\eta$, $\eta_{min}$, $\beta_{grow}$,  $\beta_{shrink}$, $\rho_0$, $\rho_1$, $\rho_2$, $w_{vc}$, $w_{nc}$, $\epsilon_{feas}$, $\epsilon_{rel}$, $N_\mathrm{it}$
\Output $\{\bar{\bm{x}}_k,\bm{u}_{\mathrm{ff},k},P_k,K_k\}_{k=1}^N$
\algcomment{Initialization}
\State Evaluate $J_\mathrm{aug}$ and $\chi$ on the initial reference using Eq. (\ref{eq:chap4_stoch_merit}) and Eq. (\ref{eq:chap4_stoch_feas})
\For{$i=1$ to $N_\mathrm{it}$}
    \algcomment{Linearization and discretization}
    \For{$k=1$ to $N-1$}
        \State $\bm{d}_k,A_k,B_k^-,B_k^+,\Sigma_{w,k}\leftarrow$ discretize about $(\bar{\bm{x}}_k,\bar{\bm{u}}_k)$ using Eq. (\ref{eq:chap4_tangent_defect}) and Eq. (\ref{eq:chap4_tangent_disc})
    \EndFor
    \algcomment{Convex subproblem}
    \State Solve Eq. (\ref{eq:chap4_ren_convex_subproblem}) in $\delta\hat{\bar{\bm{x}}},\hat{\bm{u}},\hat{\bm{\nu}},\hat{P},\hat{Y},\bm{s}$
    \algcomment{Candidate solution}
    \For{$k=1$ to $N$}
        \State $\bar{\bm{x}}_k^{+}=\delta\bar{\bm{x}}_k^\star\oplus_{G,l}\bar{\bm{x}}_k$ using Eq. (\ref{eq:chap4_candidate_reconstruction})
        \State $P_k=S_P\hat{P}_k^\star S_P^\top$, $Y_k=\hat{Y}_k^\star S_P^\top$, $K_k=Y_kP_k^{-1}$ using Eq. (\ref{eq:chap4_feedback_recovery})
    \EndFor
    \State Evaluate $J_\mathrm{aug}^\star$, $\chi^\star$, $\Delta J_\mathrm{pred}$, $\Delta J_\mathrm{act}$, $\rho$ using Eq. (\ref{eq:chap4_stoch_merit}), Eq. (\ref{eq:chap4_stoch_feas}), Eq. (\ref{eq:chap4_scp_improvements}), and Eq. (\ref{eq:chap4_scp_rho})
    \algcomment{Trust region update}
    \If{$\rho\geq\rho_0$}
        \State $\bar{\bm{x}}_k\leftarrow\bar{\bm{x}}_k^{+}$, $\bar{\bm{u}}_k\leftarrow\bm{u}_{\mathrm{ff},k}^\star$
        \State $J_\mathrm{aug}\leftarrow J_\mathrm{aug}^\star$, $\chi\leftarrow\chi^\star$
        \State Update $\eta$ using $\rho_1,\rho_2$
    \Else
        \State Shrink $\eta$
    \EndIf
    \If{Eq. (\ref{eq:chap4_feasible_candidate}) and either Eq. (\ref{eq:chap4_scp_rel_improv}) or Eq. (\ref{eq:chap4_scp_step}) holds}
        \State \textbf{break}
    \EndIf
\EndFor
\end{algorithmic}
\end{algorithm}

\section{Results}\label{sec:results}

The numerical study uses the rendezvous scenario from \cite{zhang2023stoch6DOF}. Table~\ref{tab:chap4_ren_params} summarizes the boundary conditions and physical parameters whereas Table~\ref{tab:chap4_SCvx} gives the algorithm parameters used for both the deterministic initialization and the stochastic problem. The initial and final poses are listed in the equivalent MRP/position form for readability. The final time is fixed at $t_f=200\,\mathrm{s}$. The algorithm is implemented in MATLAB using YALMIP \cite{lofberg_yalmip_2004}, and each convex subproblem is solved with MOSEK \cite{mosek_toolbox_manual}.

\begin{table}[H]
\centering
\caption{Rendezvous boundary conditions and physical parameters.}
\label{tab:chap4_ren_params}
\begin{tabularx}{0.92\textwidth}{llX}
\hline
\textbf{Parameter} & \textbf{Value} & \textbf{Description}\\
\hline
$\bm{\theta}_0$ & $[0.34,\ 0.41,\ 0.37]^\top$ & Initial MRP attitude\\
$\bm{\rho}_0$ & $[10,\ -40,\ 2]^\top\,\mathrm{m}$ & Initial relative position\\
$\bm{\omega}_0,\bm{v}_0$ & $\bm{0}$ & Initial angular velocity and relative velocity\\
$\bm{\sigma}_f$ & $[0,\ 0,\ -0.4142]^\top$ & Final MRP attitude\\
$\bm{\rho}_f$ & $[0,\ 3.7,\ 0]^\top\,\mathrm{m}$ & Final relative position\\
$\bm{\omega}_f$ & $[0,\ 0,\ 0.001]^\top\,\mathrm{rad/s}$ & Final angular velocity\\
$\bm{v}_f$ & $\bm{0}$ & Final relative velocity\\
\hline
$m$ & $961\,\mathrm{kg}$ & Chaser mass\\
$J$ & $\mathrm{diag}(2015,\ 1897,\ 1357)\,\mathrm{kg\,m^2}$ & Chaser inertia\\
$n$ & $0.001\,\mathrm{rad/s}$ & Target mean motion\\
$\tau_{\max}$ & $8.0\,\mathrm{N\,m}$ & Maximum torque magnitude\\
$F_{\max}$ & $14.0\,\mathrm{N}$ & Maximum force magnitude\\
\hline
\end{tabularx}
\end{table}

\begin{table}[H]
\centering
\caption{SCvx algorithm parameters used for deterministic initialization and stochastic optimization.}
\label{tab:chap4_SCvx}
\begin{tabularx}{0.92\textwidth}{lcc}
\hline
\textbf{Parameter} & \textbf{Deterministic initialization} & \textbf{Stochastic problem}\\
\hline
$\eta$ & $3.0{\times}10^{-2}$ & $2.5{\times}10^{-1}$\\
$\eta_{\min}$ & $1.0{\times}10^{-10}$ & $1.0{\times}10^{-10}$\\
$\eta_{\max}$ & $2.0$ & $1.0$\\
$\beta_{\mathrm{grow}}$ & $1.5$ & $1.5$\\
$\beta_{\mathrm{shrink}}$ & $0.5$ & $0.5$\\
$\rho_0$ & $0.0$ & $0.0$\\
$\rho_1$ & $0.05$ & $0.25$\\
$\rho_2$ & $0.35$ & $0.70$\\
$\epsilon_{\mathrm{rel}}$ & $1.0{\times}10^{-4}$ & $1.0{\times}10^{-3}$\\
$\epsilon_{\mathrm{feas}}$ & $1.0{\times}10^{-3}$ & $1.0{\times}10^{-3}$\\
$w_{\mathrm{vc}}$ & $1.0{\times}10^{4}$ & $1.0{\times}10^{4}$\\
Constraint slack weights & $10^{3}$ path, $10^{5}$ FOV & $10^{5}$ path/chance, $10^{3}$ control\\
\hline
\end{tabularx}
\end{table}

The target attitude (assumed constant) used to construct $C_{TH}$ is $\bm{\theta}_{TH}=-[1/3,\ 1/3,\ 1/3]^\top$. The collision and sensor geometry are:

\begin{equation}
    r_{\mathrm{KOZ}}
    =
    8\,\mathrm{m},
    \qquad
    r_x = 3\,\mathrm{m}, \, r_y = 22\,\mathrm{m} , \, r_z = 3\,\mathrm{m}
    \label{eq:ren_results_collision_values}
\end{equation}
\begin{equation}
    \bm{p}_{\mathrm{dock}}^T
    =
    [0,0,2.5]^\top,
    \qquad
    \bm{e}_{\mathrm{dock}}^T
    =
    [0,0,1]^\top,
    \qquad
    \beta_{\mathrm{dock}}
    =
    45^\circ
    \label{eq:ren_results_docking_values}
\end{equation}
\begin{equation}
    \bm{p}_{\mathrm{cam}}^C
    =
    [1,0,0]^\top,
    \qquad
    \bm{e}_{\mathrm{cam}}^C
    =
    [1,0,0]^\top,
    \qquad
    \beta_{\mathrm{FOV}}
    =
    25^\circ
    \label{eq:ren_results_fov_values}
\end{equation}

The prescribed risk level is $0.05$ for both path and control chance constraints.

The initial and terminal covariance matrices are specified in the tangent state ordering of Eq. (\ref{eq:chap4_ren_tangent_state}):

\begin{equation}
    P_0
    =
    \mathrm{blkdiag}
    \left(
        \underbrace{4{\times}10^{-6}I_3}_{\mathrm{rad}^2},\,
        \underbrace{4{\times}10^{-4}I_3}_{\mathrm{m}^2},\,
        \underbrace{4{\times}10^{-8}I_3}_{\mathrm{rad}^2\mathrm{s}^{-2}},\,
        \underbrace{4{\times}10^{-6}I_3}_{\mathrm{m}^2\mathrm{s}^{-2}}
    \right)
    \label{eq:chap4_ren_P0}
\end{equation}

\begin{equation}
    P_f
    =
    \frac{1}{2} P_0
    =
    \mathrm{blkdiag}
    \left(
        \underbrace{2{\times}10^{-6}I_3}_{\mathrm{rad}^2},\,
        \underbrace{2{\times}10^{-4}I_3}_{\mathrm{m}^2},\,
        \underbrace{2{\times}10^{-8}I_3}_{\mathrm{rad}^2\mathrm{s}^{-2}},\,
        \underbrace{2{\times}10^{-6}I_3}_{\mathrm{m}^2\mathrm{s}^{-2}}
    \right)
    \label{eq:chap4_ren_Pf}
\end{equation}

The continuous-time diffusion matrix is diagonal in the same tangent coordinates, with units per square root normalized time:

\begin{equation}
    G_c
    =
    \mathrm{diag}
    \left(
        \underbrace{2{\times}10^{-4}\bm{1}_3^\top}_{\mathrm{rad}/\sqrt{\tau}},\,
        \underbrace{5{\times}10^{-4}\bm{1}_3^\top}_{\mathrm{m}/\sqrt{\tau}},\,
        \underbrace{2{\times}10^{-5}\bm{1}_3^\top}_{\mathrm{rad}\,\mathrm{s}^{-1}/\sqrt{\tau}},\,
        \underbrace{5{\times}10^{-4}\bm{1}_3^\top}_{\mathrm{m}\,\mathrm{s}^{-1}/\sqrt{\tau}}
    \right)
    \label{eq:chap4_ren_Gc}
\end{equation}

The initial reference for the stochastic solve is generated by running the same intrinsic SCvx loop summarized in Algorithm~\ref{AlgChap4GeometricSSCVX}, but with the probabilistic components disabled: no covariance recursion is optimized, no feedback gains are synthesized, and no chance constraints are imposed. The resulting deterministic trajectory is then used to initialize the full intrinsic stochastic SCvx method with $N=60$ nodes. Monte Carlo validation uses $200$ nonlinear stochastic trajectories sampled from the initial covariance and propagated on a finer grid. Two closed loop cases are compared: the covariance steering controller produced by the stochastic optimization, from here on referred to as \textit{isSCvx}, and a feedback linearization baseline tracking the deterministic reference trajectory, referred to as \textit{feedbacklin} and implemented as in~\cite{zhang2023stoch6DOF}.

\begin{figure}[H]
\centering
\includegraphics[width=0.6\textwidth]{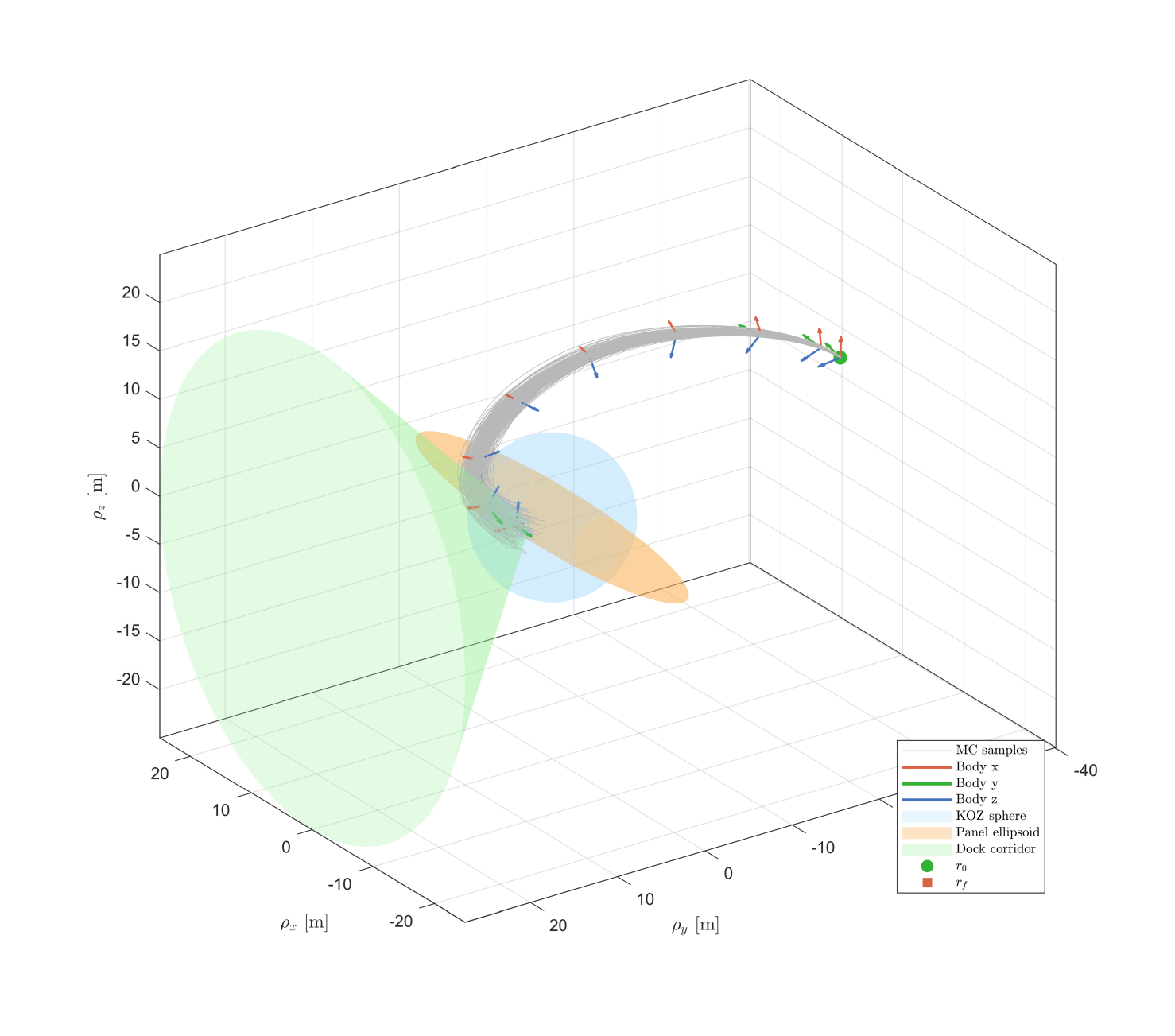}
\caption{Open loop Monte Carlo rendezvous trajectories obtained by propagating the optimized feedforward trajectory through the nonlinear stochastic dynamics without feedback correction.}
\label{fig:ren_trajectory_ol}
\end{figure}

Figure~\ref{fig:ren_trajectory_ol} shows the open loop propagated trajectories. All Monte Carlo samples are initialized from the prescribed initial covariance and then propagated with the same optimized feedforward input, so the spread visible in the plot is the direct consequence of the initial dispersion and process noise acting through the nonlinear SE(3) dynamics. The nominal trajectory reaches the docking region, but the samples are not actively corrected as they deviate from the reference. This makes the open loop plot useful as a baseline: it shows that nominal feasibility of the deterministic trajectory is not enough to characterize the behavior of the stochastic trajectory distribution.

\begin{figure}[H]
\centering
\begin{subfigure}{0.49\textwidth}
    \centering
    \includegraphics[width=\textwidth]{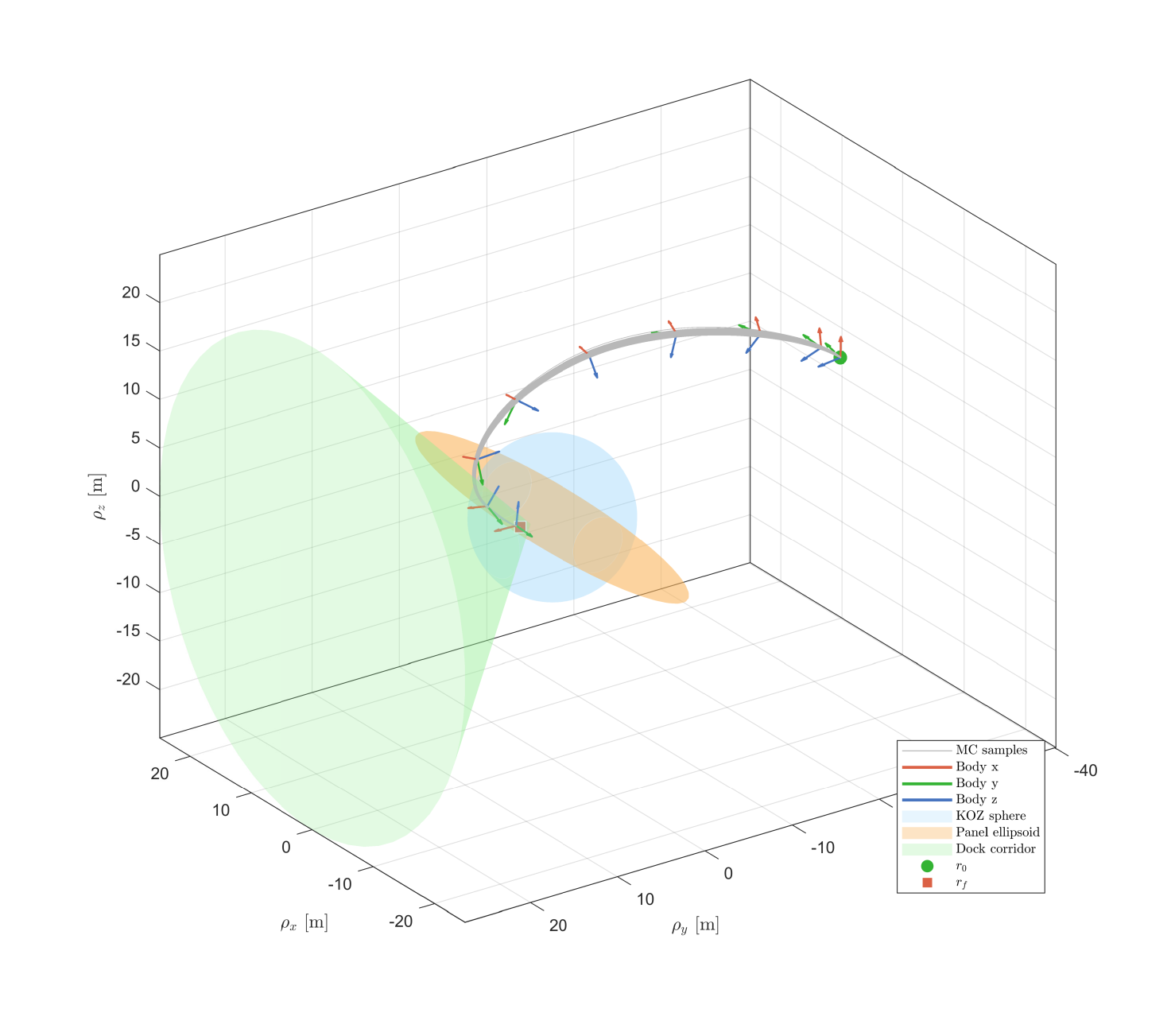}
    \caption{Intrinsic stochastic SCvx.}
    \label{fig:ren_trajectory}
\end{subfigure}
\hfill
\begin{subfigure}{0.49\textwidth}
    \centering
    \includegraphics[width=\textwidth]{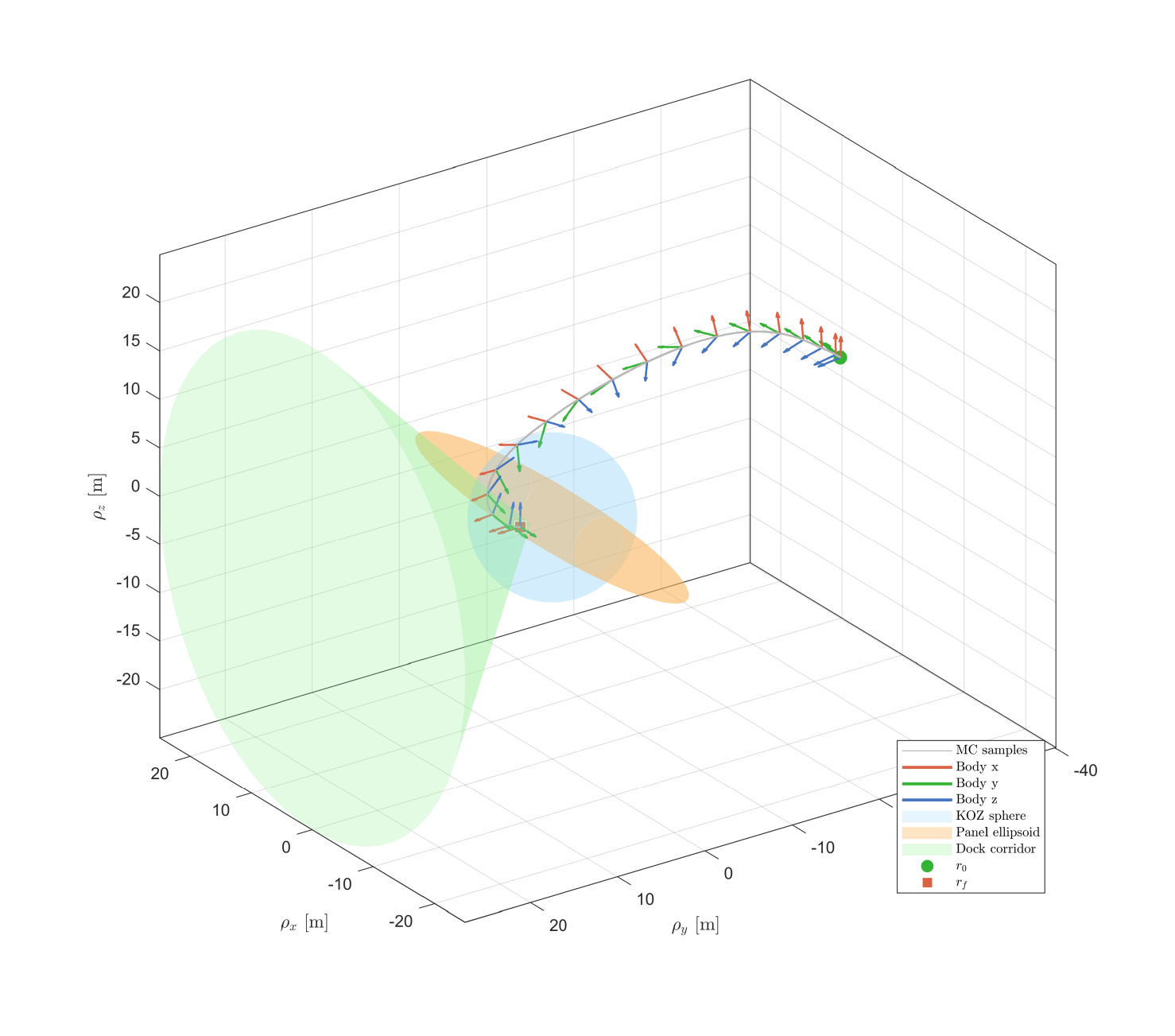}
    \caption{Feedback linearization.}
    \label{fig:ren_trajectory_feedbacklin}
\end{subfigure}
\caption{Closed loop Monte Carlo rendezvous trajectories for the intrinsic stochastic SCvx feedback policy and the feedback linearization controller.}
\label{fig:ren_trajectory_cl}
\end{figure}

Figure~\ref{fig:ren_trajectory_cl} compares the two closed loop alternatives. Both controllers use feedback to keep the samples close to the reference approach, but they are designed with different objectives. The \textit{feedbacklin} controller tracks the deterministic trajectory after it has been generated, whereas the \textit{isSCvx} controller is synthesized together with the nominal trajectory and covariance sequence. Therefore, the 3D trajectories should be interpreted as a qualitative first indication of closed loop behavior: the important distinction is not only whether the samples remain visually close to the reference, but whether their dispersion is shaped consistently with the terminal covariance requirement and the chance constrained collision and sensor pointing metrics examined in the following plots.

\begin{figure}[H]
\centering
\begin{subfigure}{\textwidth}
    \centering
    \includegraphics[width=0.95\textwidth]{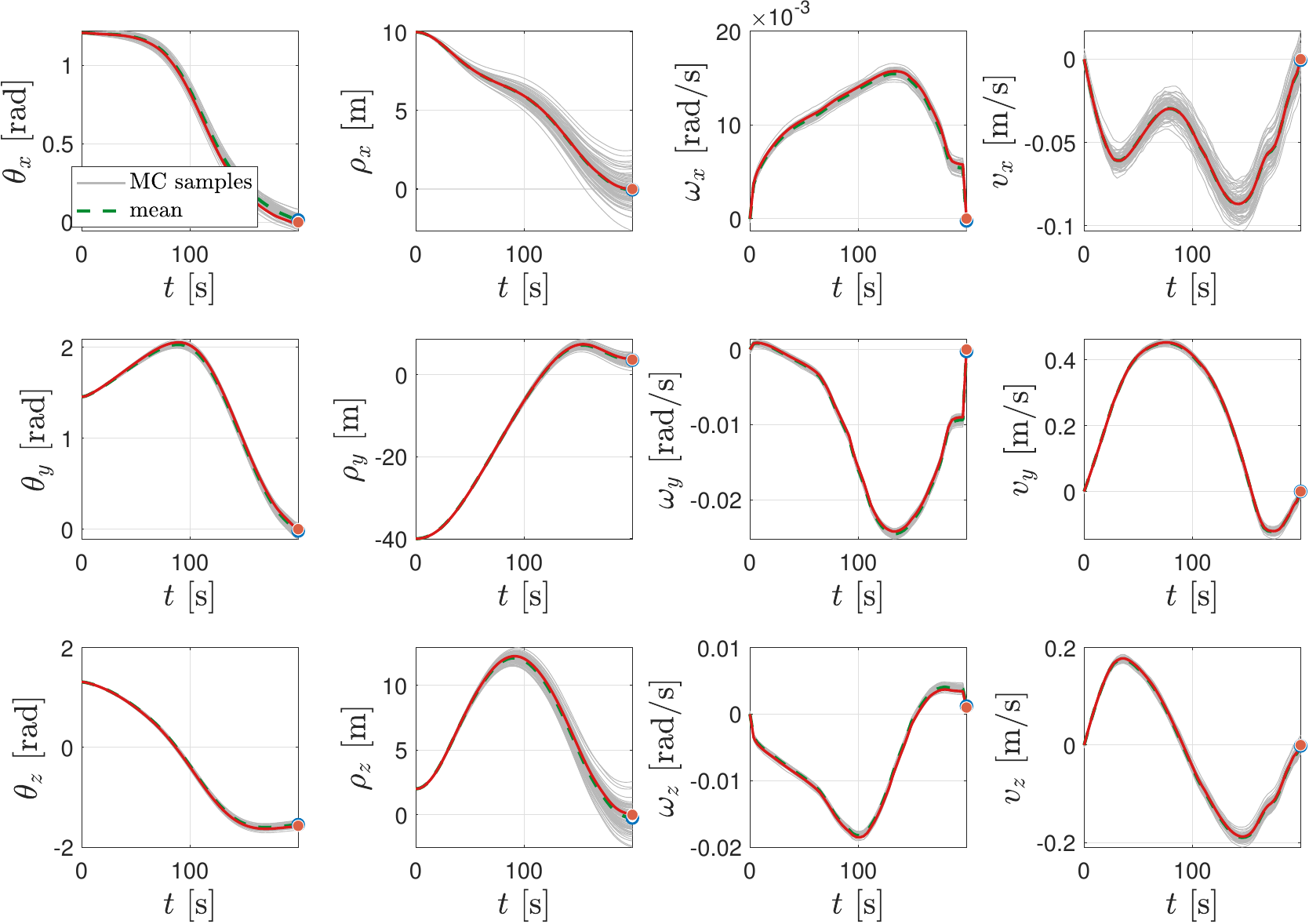}
    \caption{State histories.}
    \label{fig:ren_ol_states}
\end{subfigure}
\\[1em]
\begin{subfigure}{0.48\textwidth}
    \centering
    \includegraphics[width=\textwidth]{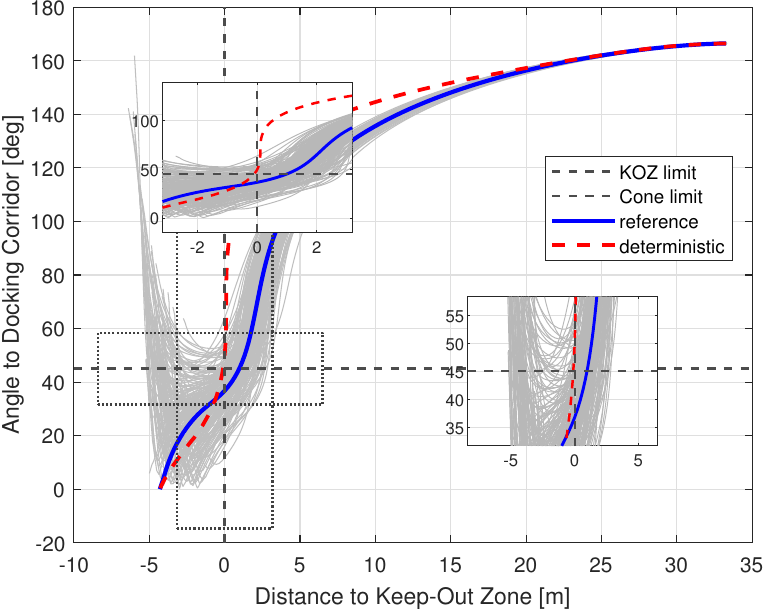}
    \caption{Collision avoidance constraints.}
    \label{fig:ren_ol_collision}
\end{subfigure}
\hfill
\begin{subfigure}{0.48\textwidth}
    \centering
    \includegraphics[width=\textwidth]{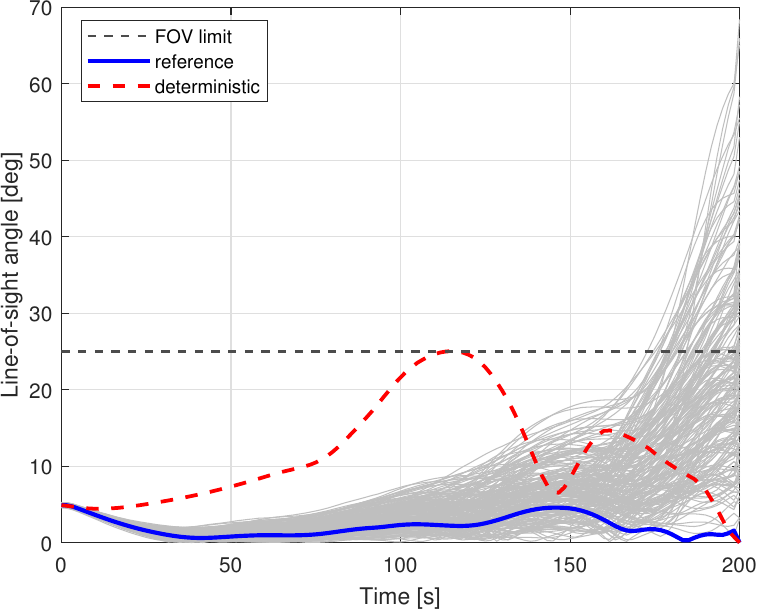}
    \caption{Camera field of view constraint.}
    \label{fig:ren_ol_fov}
\end{subfigure}
\caption{Open loop Monte Carlo propagation of the optimized feedforward trajectory.}
\label{fig:ren_open_loop}
\end{figure}

Figure~\ref{fig:ren_open_loop} shows the open loop propagated states (Fig.~\ref{fig:ren_ol_states}), the collision constraints (Fig.~\ref{fig:ren_ol_collision}), and FOV constraint (Fig.~\ref{fig:ren_ol_fov}) of the optimized feedforward trajectory through the nonlinear stochastic dynamics. In the state-history plots, the SE(3) pose is displayed as equivalent MRP attitude $\bm{\sigma}$ and Hill-frame position $\bm{\rho}^H$ for readability; these plotted attitude components are not the tangent covariance coordinates. The feedforward solution reaches the desired docking corridor, but the Monte Carlo envelope spreads around the nominal path. This open loop behavior is a key reason for optimizing the covariance and feedback gain together with the nominal trajectory: constraint satisfaction must be achieved for the controlled distribution, not merely for the deterministic reference. The red dashed line in the collision and FOV constraint plots corresponds to the deterministic solution, which remains very close to the constraint boundaries.

\begin{figure}[H]
\centering
\begin{subfigure}{0.48\textwidth}
    \centering
    \includegraphics[width=\textwidth]{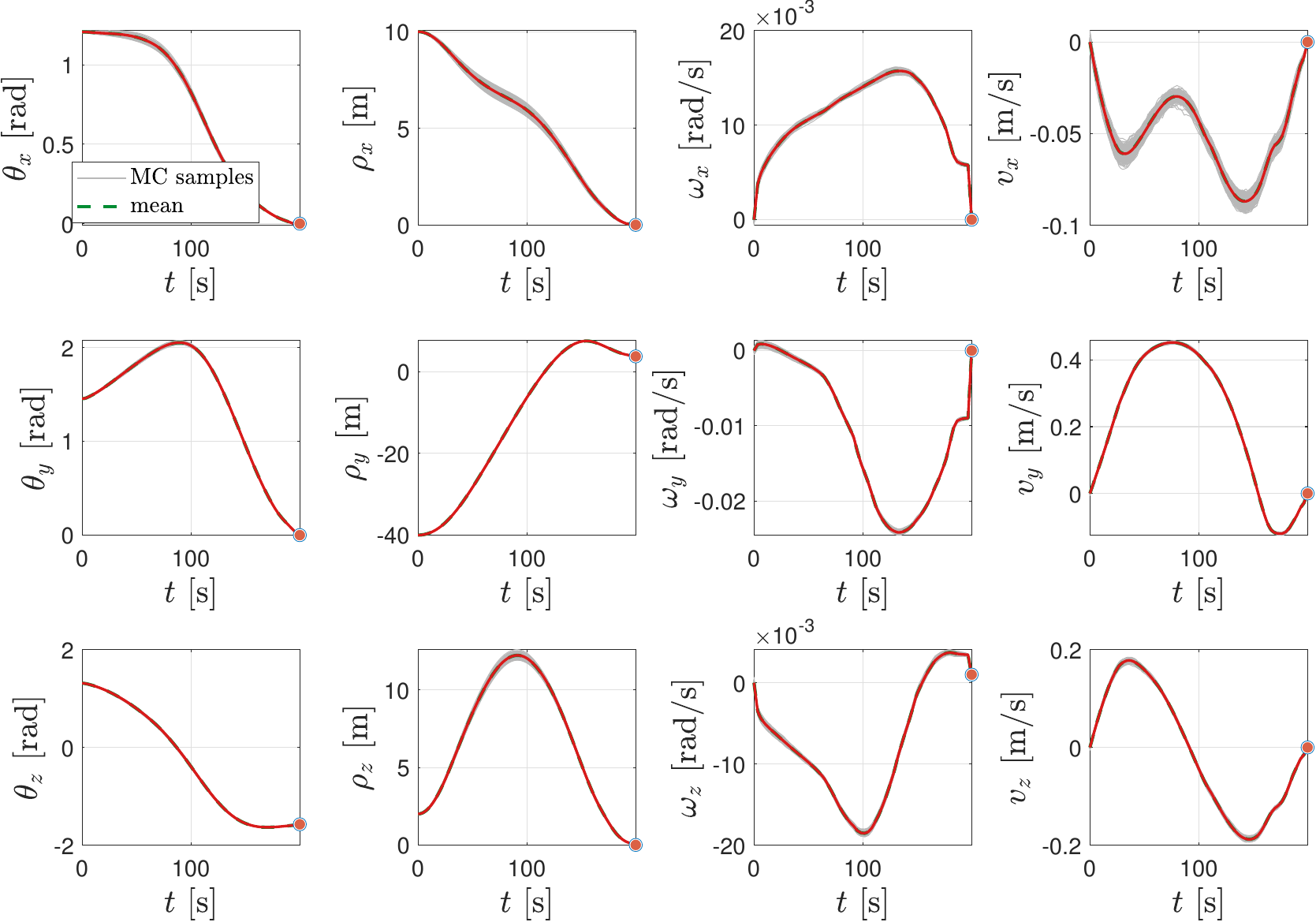}
    \caption{State histories.}
    \label{fig:ren_states}
\end{subfigure}
\hfill
\begin{subfigure}{0.48\textwidth}
    \centering
    \includegraphics[width=\textwidth]{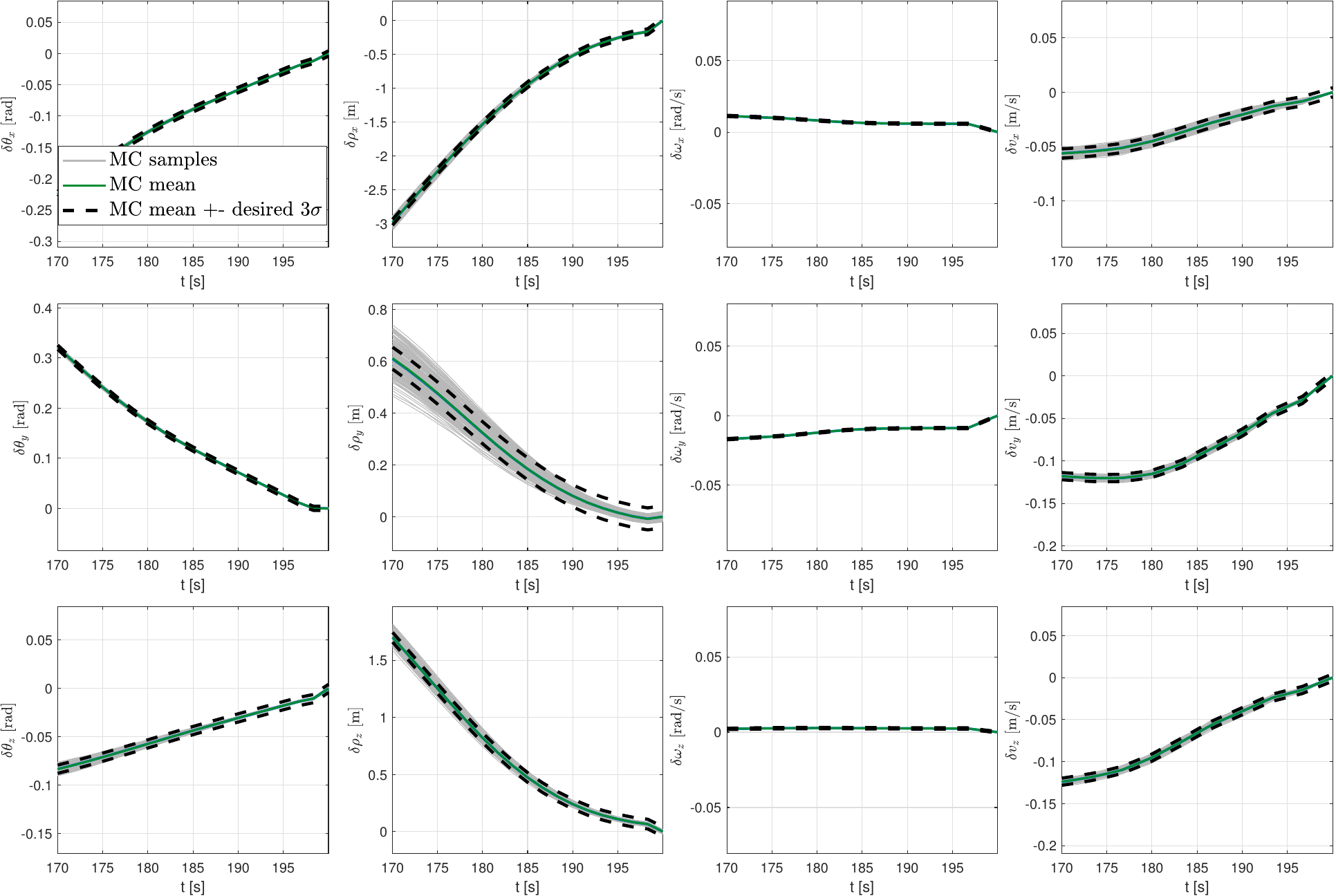}
    \caption{Terminal covariance.}
    \label{fig:ren_final_cov}
\end{subfigure}
\caption{Closed loop rendezvous state histories and terminal covariance comparison for the intrinsic stochastic SCvx feedback policy.}
\label{fig:ren_states_cov}
\end{figure}

Figure~\ref{fig:ren_states_cov} reports the closed loop state histories (Fig. \ref{fig:ren_states}) and the terminal covariance comparison (Fig. \ref{fig:ren_final_cov}) for the \textit{isSCvx} controller. As in Fig.~\ref{fig:ren_ol_states}, the state histories display the pose using equivalent MRP attitude $\bm{\sigma}$ and Hill-frame position $\bm{\rho}^H$, together with angular and translational velocity components. By contrast, the terminal covariance focus plot is expressed in the local tangent error coordinates used by the optimizer: the attitude entries are principal rotation-vector errors $\delta\bm{\theta}$ obtained from the SO(3) logarithm of the attitude error, not absolute MRP states. The state histories show the Monte Carlo samples and their mean, while the terminal covariance focus plot shows that the final dispersions are successfully steered to the desired values.

\begin{figure}[H]
\centering
\begin{subfigure}{0.48\textwidth}
    \centering
    \includegraphics[width=\textwidth]{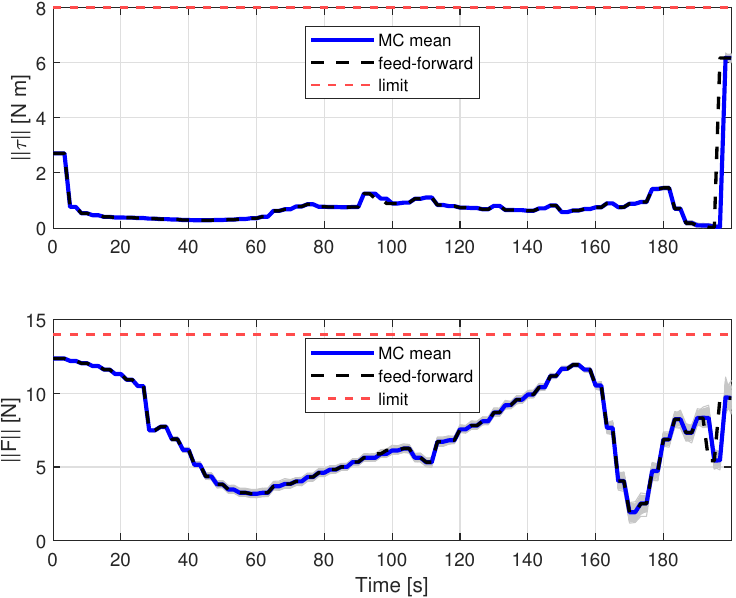}
    \caption{Intrinsic stochastic SCvx.}
\end{subfigure}
\hfill
\begin{subfigure}{0.48\textwidth}
    \centering
    \includegraphics[width=\textwidth]{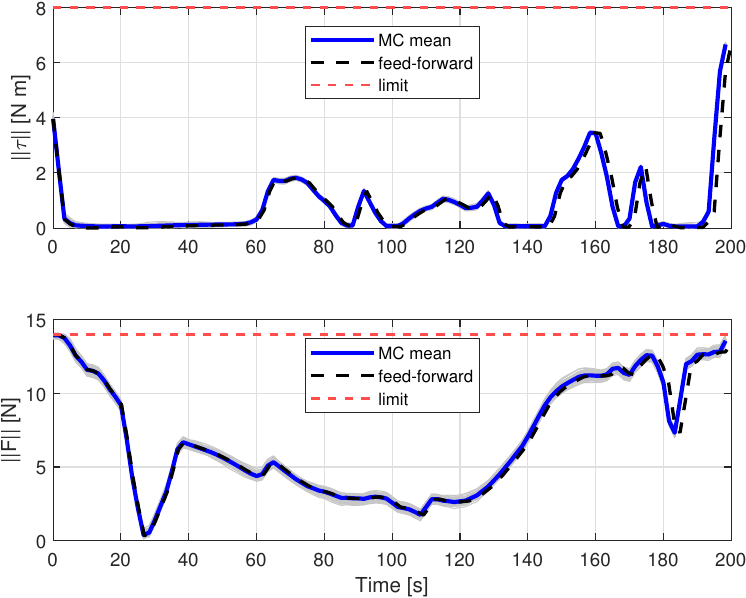}
    \caption{Feedback linearization baseline.}
\end{subfigure}
\caption{Closed loop control histories over the Monte Carlo trials. Force and torque magnitudes are shown against their admissible bounds.}
\label{fig:ren_controls}
\end{figure}

Figure~\ref{fig:ren_controls} compares the control histories. The stochastic optimization includes chance constraints on both force and torque magnitudes, so the commanded feedback effort is regularized by the same risk allocation used in the trajectory design. The baseline controller uses the same physical actuator limits for evaluation, but applies them after the tracking command is computed rather than optimizing the feedback law with respect to probabilistic control bounds, resulting in the thrust magnitude being very close to the maximum value at the beginning and end of the rendezvous.

\begin{figure}[H]
\centering
\begin{subfigure}{0.48\textwidth}
    \centering
    \includegraphics[width=\textwidth]{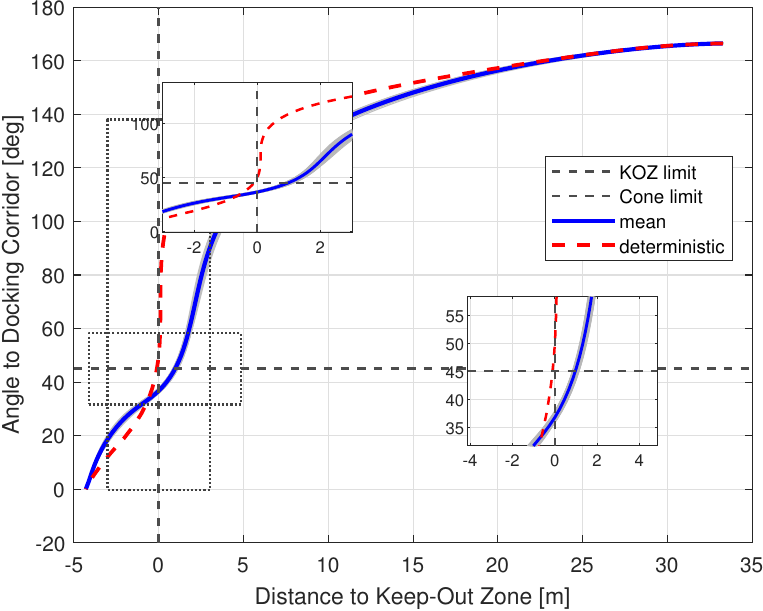}
    \caption{Intrinsic stochastic SCvx.}
\end{subfigure}
\hfill
\begin{subfigure}{0.48\textwidth}
    \centering
    \includegraphics[width=\textwidth]{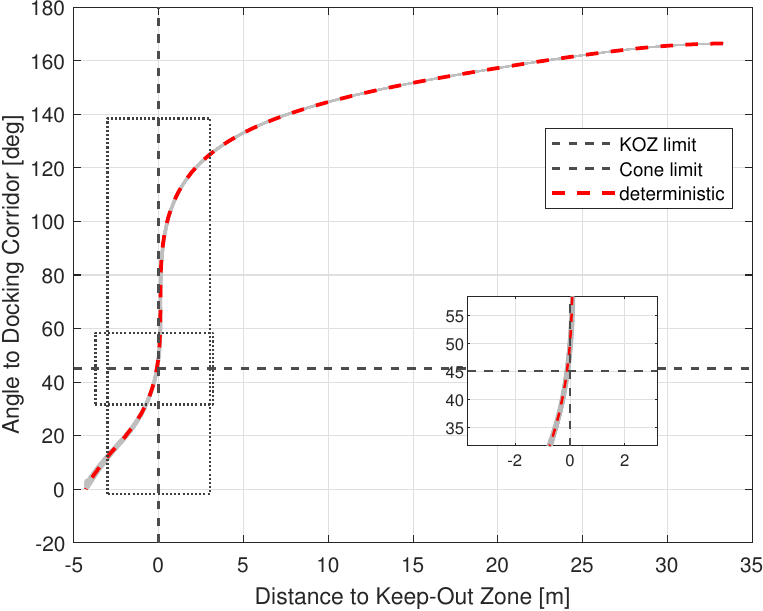}
    \caption{Feedback linearization baseline.}
\end{subfigure}
\caption{Closed loop collision avoidance metrics. The plots show Monte Carlo clearance relative to the keep out and solar panel constraints during the approach.}
\label{fig:ren_collision}
\end{figure}

\begin{figure}[H]
\centering
\begin{subfigure}{0.48\textwidth}
    \centering
    \includegraphics[width=\textwidth]{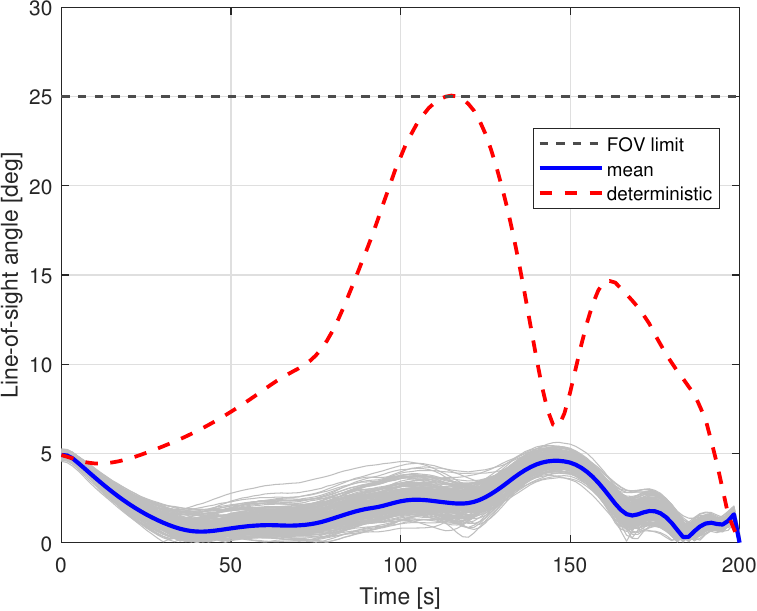}
    \caption{Intrinsic stochastic SCvx.}
\end{subfigure}
\hfill
\begin{subfigure}{0.48\textwidth}
    \centering
    \includegraphics[width=\textwidth]{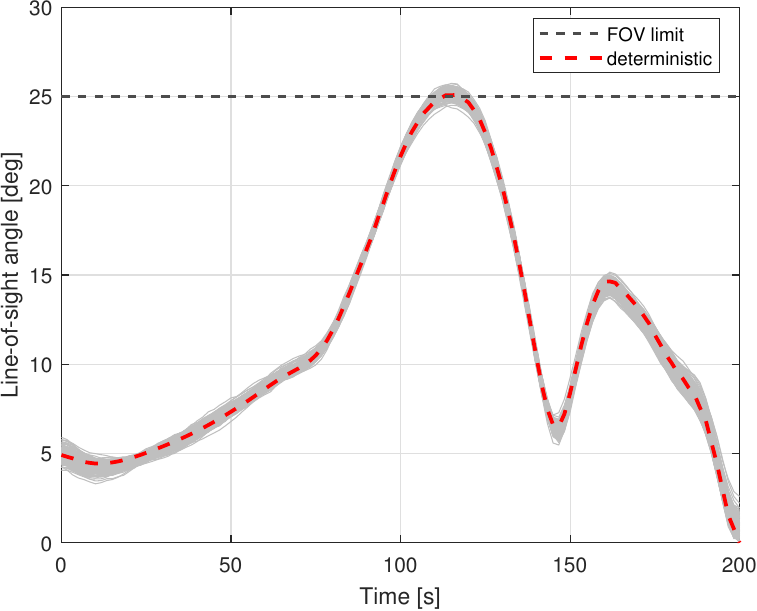}
    \caption{Feedback linearization baseline.}
\end{subfigure}
\caption{Closed loop camera field of view angle over the Monte Carlo trials. The horizontal limit corresponds to the prescribed field of view half angle.}
\label{fig:ren_fov}
\end{figure}

Figures~\ref{fig:ren_collision} and \ref{fig:ren_fov} show the coupled position attitude constraints that motivate a pose based stochastic formulation. Collision avoidance depends on the relative position with respect to the target geometry, while the field of view constraint depends simultaneously on relative position and chaser attitude. The intrinsic stochastic SCvx solution plans the nominal pose, tangent covariance, and feedback response jointly, so the Monte Carlo dispersion is shaped with respect to these nonlinear safety metrics. The comparison with feedback linearization illustrates that tracking a deterministic trajectory is not equivalent to solving the chance constrained trajectory design problem: even when the nominal path is feasible, the closed loop distribution can interact with the nonlinear constraints in a qualitatively different way.

\begin{figure}[H]
\centering
\includegraphics[width=\textwidth]{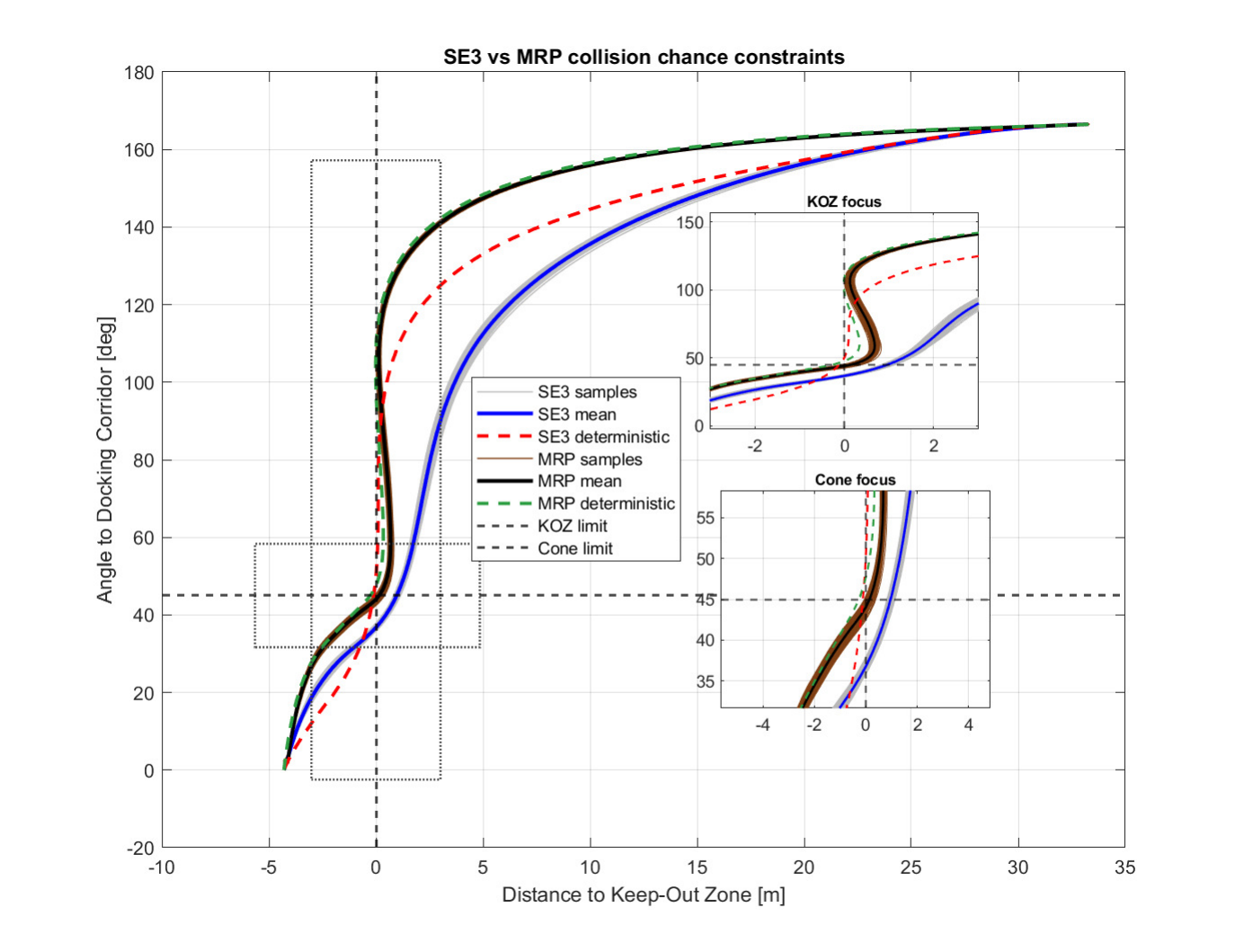}
\caption{Collision avoidance comparison between the intrinsic SE(3) solution and the MRP-position solution for the rendezvous scenario.}
\label{fig:ren_collision_mrp_comp}
\end{figure}

\begin{figure}[H]
\centering
\includegraphics[width=\textwidth]{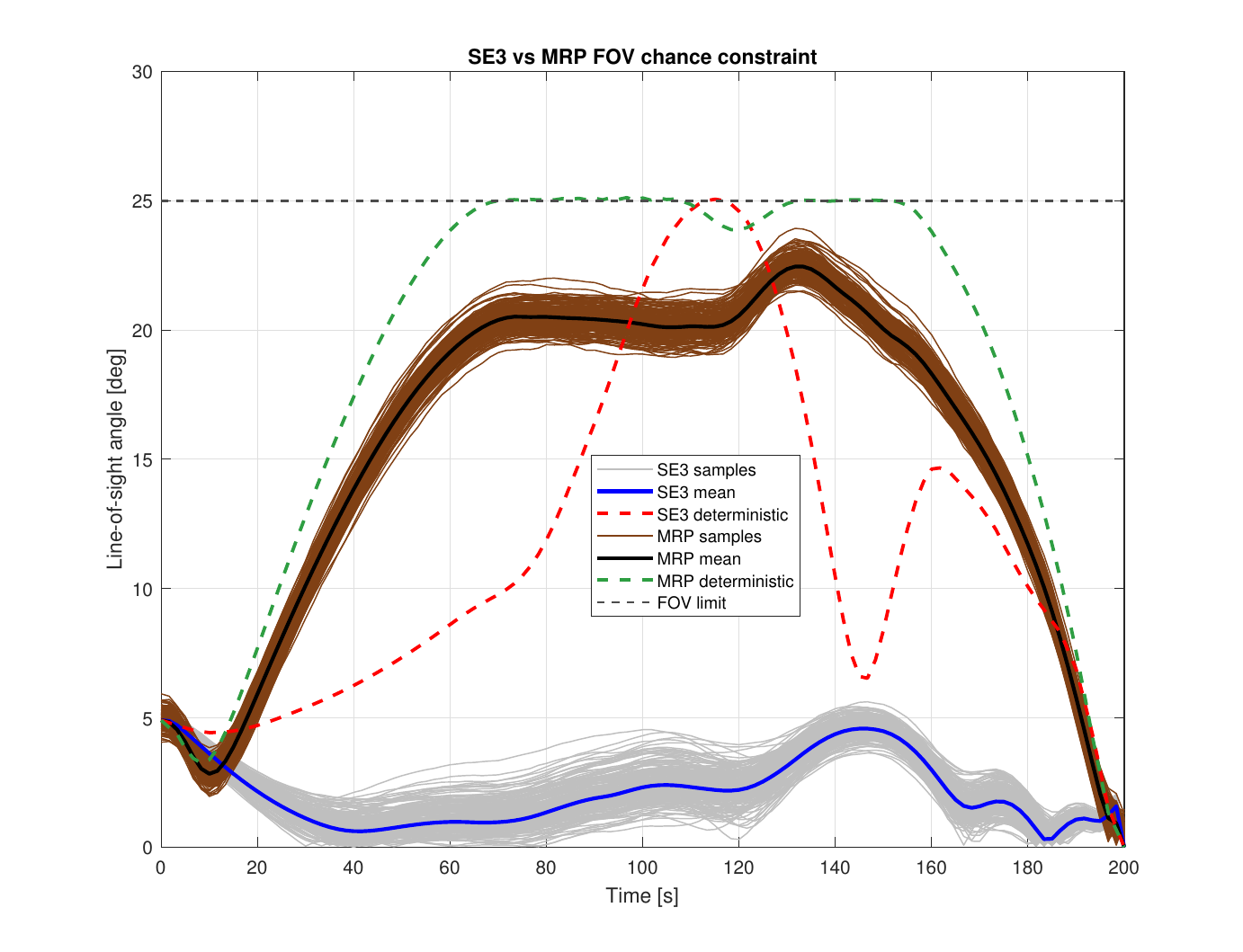}
\caption{Camera field of view comparison between the intrinsic SE(3) solution and the MRP-position solution for the rendezvous scenario.}
\label{fig:ren_fov_mrp_comp}
\end{figure}

Finally, Figures~\ref{fig:ren_collision_mrp_comp} and~\ref{fig:ren_fov_mrp_comp} further compare the intrinsic SE(3) formulation with a stochastic trajectory optimization solution that represents attitude with modified Rodrigues parameters and position in Euclidean coordinates as in \cite{zhang2023stoch6DOF}. The two formulations use the same physical rendezvous scenario and safety constraints, but differ in how pose perturbations, covariance propagation, and feedback corrections are modeled. In both the collision avoidance and field of view metrics, the MRP-position solution remains noticeably closer to the constraint boundaries than the intrinsic SE(3) solution. This behavior is consistent with the fact that the MRP-position formulation treats attitude and position uncertainties as separate Euclidean components, whereas the intrinsic formulation propagates and controls the coupled pose uncertainty on SE(3): neglecting that coupling limits how effectively the optimizer can shape dispersion in the nonlinear position-attitude constraints.

\section{Conclusion}\label{sec6}
This paper presented an intrinsic stochastic successive convexification method for chance constrained six degree of freedom trajectory optimization on SE(3). The central idea is to keep the accepted pose trajectory on the Matrix Lie Group while expressing stochastic perturbations, covariance propagation, feedback synthesis, and chance constraint transcriptions in local tangent coordinates of the Lie algebra se(3) using the concept of concentrated probability distributions. 

The rendezvous example showed how the method can handle coupled position-attitude safety constraints and probabilistic control magnitude limits. Compared with tracking a deterministic reference using a feedback linearization controller, the proposed approach directly optimizes the nominal trajectory, covariance sequence, and feedback gains. The resulting closed loop Monte Carlo simulations demonstrate improved satisfaction of the probabilistic path and control constraints, confirming that designing the trajectory and feedback law together is advantageous when stochastic dispersion interacts with nonlinear pose-dependent constraints.

Future work will extend the implementation to include higher fidelity relative dynamics, navigation uncertainty, and actuation errors. Another natural extension is to carry out the same intrinsic stochastic SCvx derivation using unit dual quaternions, which share the same se(3) tangent representation while providing an alternative algebraic realization of rigid body pose.

\bibliography{sample}

\end{document}